\documentclass[manuscript]{aastex}

\shorttitle{Evolution of Coronal Plasma Parameters at Solar Minimum}
\shortauthors{Strachan, et al.}

\begin{document}

%% LaTeX will automatically break titles if they run longer than
%% one line. However, you may use \\ to force a line break if
%% you desire.

\title{THE EVOLUTION OF PLASMA PARAMETERS ON A \\
CORONAL SOURCE SURFACE AT $2.3~R_{\sun}$ DURING SOLAR MINIMUM }

%% Use \author, \affil, and the \and command to format
%% author and affiliation information.
%% Note that \email has replaced the old \authoremail command
%% from AASTeX v4.0. You can use \email to mark an email address
%% anywhere in the paper, not just in the front matter.
%% As in the title, use \\ to force line breaks.

%%\author{S. Djorgovski\altaffilmark{1,2,3} and Ivan R. King\altaffilmark{1}}
%%\affil{Astronomy Department, University of California,
%%    Berkeley, CA 94720}

\author{L. Strachan, A. V. Panasyuk, J. L. Kohl}
\affil{Harvard-Smithsonian Center for Astrophysics, 60 Garden St., Cambridge, 
       MA 02138, USA}
\email{lstrachan@cfa.harvard.edu}

\and

\author{P. Lamy}
\affil{Laboratoire d'Astrophysique de Marseille, UMR6110 CNRS/Universit\'{e} de Provence, \\
38 rue Fr\'{e}d\'{e}ric Joliot-Curie, 13388 Marseille Cedex 13, France}

%=============================================
%% Notice that each of these authors has alternate affiliations, which
%% are identified by the \altaffilmark after each name.  Specify alternate
%% affiliation information with \altaffiltext, with one command per each
%% affiliation.

%\altaffiltext{1}{Visiting Astronomer, Cerro Tololo Inter-American Observatory.
%CTIO is operated by AURA, Inc.\ under contract to the National Science
%Foundation.}
%==============================================

\begin{abstract}

%%%%%%%%%%%%%%%%%%%%%%%%%%%%%%%%%%%
%% The final abstract should be a ***distillation of the entire paper***
%% (not just a table of contents)
%%
%% Include all of the following:
%% - Present the main results 
%% - Discuss how the work was accomplished
%% - Relate to previous work
%% - Discuss why the work is important
%%
%%%%%%%%%%%%%%%%%%%%%%%%%%%%%%%%%%%

We analyze data from the {\it Solar and Heliospheric Observatory} to produce global maps of coronal outflow velocities and densities in the regions where the solar wind is undergoing acceleration. The maps use UV and white light coronal data obtained from the  {Ultraviolet Coronagraph Spectrometer} and the {Large Angle Spectroscopic Coronagraph}, respectively, and a Doppler dimming analysis to determine the mean outflow velocities. The outflow velocities are defined on a sphere at $2.3~ R_{\sun}$ from Sun-center and are organized by Carrington Rotations during the solar minimum period at the start of solar cycle 23.  We use the outflow velocity and density maps to show that while the solar minimum corona is relatively stable during its early stages, the shrinkage of the north polar hole in the later stages leads to changes in both the global areal expansion of the coronal hole and the derived internal flux tube expansion factors of the solar wind.  The polar hole areal expansion factor and the flux tube expansion factors (between the coronal base and $2.3~ R_{\sun}$) start out as super-radial but then they become more nearly radial as the corona progresses away from solar minimum.  The results also support the idea that the largest flux tube expansion factors are located near the coronal hole/streamer interface, at least during the deepest part of the solar minimum period.

\end{abstract}

%% Keywords should appear after the \end{abstract} command. The uncommented
%% example has been keyed in ApJ style. See the instructions to authors
%% for the journal to which you are submitting your paper to determine
%% what keyword punctuation is appropriate.


%% From the front matter, we move on to the body of the paper.
%==========================================
\section{INTRODUCTION}

The ambient solar wind is known to have essentially two states: a fast and a slow wind component   \citep{Hund72, McComas00}. The fast solar wind escapes from open magnetic field regions found in coronal holes \citep[][]{ktr73} but the slow solar wind probably has multiple sources.  The primary source of the slow solar wind seems to be near coronal streamers, i.e., from the streamer tips or from streamer/coronal hole boundaries \citep[e.g.,][]{fel81,wom97, hab97, Wang1998, str00, AAD05}.  Other candidates for source regions of the slow wind are active regions and small equatorial coronal holes, \citep[e.g., see][]{Kojima99,LNZ04,MCK04,WH05,fatz06}.  Part of the uncertainty in identifying the origin of the solar wind (particularly the slow wind) is that most measurements of the wind speed are made at heights far above the source regions.  Mapping the streams back to the Sun is not always straightforward due to differences in the speed of the wind and the interactions between fast and slow wind as the wind flows away from the Sun.  Early attempts to measure the solar wind speeds near the Sun relied on {in situ} velocity measurements that were extrapolated back to the Sun using potential field magnetic models \citep[e.g., see][and references therein]{AP00}. These extrapolations also depend on assumptions made about the expansion factors for the diverging coronal magnetic field, which serves as streamlines for the outflowing plasma.

Solar wind velocity maps have been made for more than two decades using radio interplanetary scintillation (IPS) techniques. For example, \citet {Kojima87,Kojima98,Breen00,TKF10} have made maps of outflow speeds on a source surface by interpreting the scintillation pattern from distant radio sources such as quasars.  The scintillations carry information about the speeds of small scale structures moving away from the Sun in the solar wind.  There are several different tomographic methods that can be used to extract the three dimensional structure from the scintillation data \citep[e.g., see the reviews by][]{Kojima07, Jackson11}. However creating near-Sun maps from these data must be made carefully, since the minimum distance of the source-receiver lines of sight for these measurements are typically at $\sim 20~R_{\sun}$ from the Sun (depending on frequency).  In order to make the velocity maps at $\sim 2.5~R_{\sun}$, some type of extrapolation must be used, which can have its own uncertainties as already stated.  Despite these limitations, an important contribution of the IPS measurements is that the velocity maps produced from this technique cover all heliographic latitudes.  These results are the most similar to the work presented in this paper and thus invites a comparison with our results which will be described below.

Another method for determining outflow velocities near the Sun is to track brightness inhomogeneities (or
blobs) that appear in white light coronagraph images. This technique has been used successfully with the
Large Angle Spectroscopic Coronagraph or LASCO \citep[][]{bru95} and more recently with the SECCHI suite of coronagraphs and
heliospheric imagers \citep[][]{howard08} on the Solar Terrestrial Relations Observatory mission.  If the blobs are
considered to be tracers of the solar wind \citep{she97}, then their outflow speeds can be used to measure
the bulk outflow velocity (projected onto the plane of the sky).  Movies of the difference images of
successive frames show structures moving out from the LASCO-C2 inner field of view at $\sim 2~R_{\sun}$ to
the LASCO-C3 outer field of view at $\sim 30~R_{\sun}$.  \citet{she97} used scatter plots of many blobs to
show that outflow speeds start out near zero at $\sim 3~R_{\sun}$, rise rapidly with height, and then
gradually increase to $300 ~km~s^{-1}$ at and beyond $\sim 20~R_{\sun}$.  Because the blobs appear to emerge
from streamer tips and the termination speeds are similar to that expected for the slow solar wind, the data
suggest that streamers are likely sources of the slow speed wind.  If the blob material does indeed form
part of the solar wind, it is likely that it contains mostly material from the streamer legs, i.e., the
bright regions in \ion{O}{6} emission on either side of the central core.   This is suggested by the work of \citet{ray97} which shows that the minor ion abundances in the streamer legs are similar to those measured in the slow solar wind at 1~AU.  However, there is still some debate about whether the blob material comes from streamer evaporation at the tips or from foot point reconnections at the streamer base \citep[see][]{Wang00,Jones09}.   While feature tracking works well for determining outflows from streamers, it has been less successful in tracking features in the much dimmer coronal holes where the fast solar wind originates.  

With this new work we produce a two dimensional map of coronal outflow velocities covering the entire Sun at a fixed height by using a Doppler dimming analysis of coronal emission lines \citep[e.g., see][]{wkwm82,noc87,xli98,cran99}.   The present work builds on the techniques developed for providing outflow velocity measurements as a function of height and latitude that were described in our previous study \citep[see][]{str02}. The advantages of using the Doppler dimming technique is that the outflow velocities can be determined much closer to the Sun. In fact, the most sensitive height ranges for observing coronal UV emission lines with the Ultraviolet Coronagraph Spectrometer (UVCS) \citep{kohl95} is in the heliocentric height range $r ~=~ 1.5 - 3.5~R_{\sun}$. This is the region where UVCS observations have shown that much of the coronal heating and solar wind acceleration takes place \citep{kohl98,cran99}.  The outflow velocities are defined on a reference sphere at $2.3~R_{\sun}$ which we will refer to as a source surface.  The choice for the reference height was made so that the outflow velocity measurements would be as close as practical to the traditional magnetic field source surface at $2.5~R_{\sun}$ \citep[e.g., see][]{Schatten1969}.  The maps, which are organized by Carrington rotations (CRs), were not constructed at exactly $2.5~R_{\sun}$ since there are significant gaps in the UVCS synoptic observations at this height. 

The solar minimum period is used as a starting point for the velocity maps since the corona is in its simplest state at that time in the solar cycle.  It is also an ideal time for comparing conditions in the corona with predictions from magnetohydrodynamic (MHD) models.  For the purposes of this study, we define the time between 1996 May and 1997 January as the period that is characteristic of low solar activity (the `solar minimum period').  \citet{HW99} considered  the same time period in their analysis to define the minimum of cycle 23.  The first date is the time when the smoothed monthly sunspot number was at a minimum and the second date is when the sunspot number in the old cycle was equal to number in the new cycle.  By using seven different solar parameters they determined that the actual minimum of solar activity for cycle 23 occurred in 1996 September.  The maps produced in the present paper cover the period from 1996 May to 1998 June, which includes this period of low solar activity as well as part of the rising phase which started in 1997.

This paper describes the evolution of the outflow velocity and density structures in the corona with a discussion of a few derived parameters such as the particle flux and two types of solar wind expansion factors.  The paper is organized as follows:  In section \ref{empmod}, we describe the line of sight coronal model that is used to produce the coronal outflow velocities.  Maps of outflow velocity and electron density and their variations over the solar minimum period are presented in section \ref{nvmaps}.  We also compare our outflow velocity maps with those produced by IPS observations at the end of this section. Section \ref{discuss} contains a discussion of the latitudinal and temporal variation of the derived solar wind particle flux and flux tube expansion factor.  A summary of the results and a discussion of future work are provided in section \ref{concl}.   
%==========================================
%% You can use LaTeX's \ref and \label commands to keep track of
%% cross-references to sections, equations, tables, and figures.
\section{CORONAL EMISSION LINE MODEL} \label{empmod}

The coronal outflow velocities for  O$^{5+}$ ions are determined from a self-consistent model of the
observed \ion{O}{6} 103.2 and 103.7 nm intensities.  The method uses a spectral line synthesis coronal code called {\it CORPRO} to compute modeled
\ion{O}{6} 103.2 and 103.7 nm spectral emissivities, given empirical constraints for the incident \ion{O}{6} radiation from the solar disk, the coronal electron densities, and the kinetic temperatures (which includes both thermal and non-thermal motions) for the electrons and ions. The code is similar to previous codes that have been used in the past \citep[e.g.,][]{wkwm82, xli98, cran99, nmac99, str00, akinari07}; however, upgrades were required to produce the outflow velocity maps.  These include a complete rewrite of the original code so that it runs more efficiently and the construction of new algorithms for sorting the data and handling the two dimensional aspects of the inputs and outputs.

In principle, coronal outflow velocities are computed by finding the velocities which produce modeled intensities and line widths that are consistent with the observed resonantly scattered profiles. However, for the \ion{O}{6} doublet, there is a collisional component to each line which complicates the analysis. Instead we use the \ion{O}{6} 103.2 nm and 103.7 nm intensity ratios to isolate the resonantly scattered component which is sensitive to the outflow velocities.  As described below, the model intensity ratios are derived from the line-of-sight integrated emissivities for each line.  Another advantage of using the line ratios is that the absolute abundance for oxygen is not needed since this quantity cancels out when the ratio is performed.

Following \citet{str00},
the coronal emissivities for the \ion{O}{6} lines, where the
subscript $i=1$ or 2 for the 103.2 nm or 103.7 nm line,
can be expressed as the sum of a collisional component and a resonantly
scattered component:
$ {E_i} ~=~ {E_i}^{col} ~+~ {E_i}^{res} $. The expressions
for the collisional and resonantly scattered components, in
units of photons~${\rm s^{-1} cm^{-3} sr^{-1}}$, are given below.  For the collisional component, ${E_i}^{col}$, we have:

\begin{equation} \label{eqn_col}
{E_i}^{col} ~=~{\int_{- \infty}^{\infty}
\frac{1}{4 \pi} N_1({\rm O^{5+}}) N_e q_{col} 
\phi_c ( \lambda - \lambda _i ) ~d \lambda },
\end{equation}

\noindent
where $N_1$(O$^{5+}$), is the ground state number density                 
for O$^{5+}$, $N_e$ is the electron density, $q_{col}(T_e)$
is the collisional excitation rate as a function
of electron temperature, $T_e$, and
$\phi_c ( \lambda - \lambda _i )$ is the coronal line profile, with the line center
wavelength specified by $\lambda _i)$.  For the resonantly scattered component, ${E_i}^{res}$, we have:

\begin{eqnarray} \label{eqn_res}
{E_i}^{res} & = & {\int_{- \infty}^{\infty} \int_{- \infty}^{\infty}
 \int_{\Omega}} ~C~ N_1({\rm O^{5+}})
\phi_r ( \lambda - \lambda _i ) \nonumber \\
 & \times & I_D ( \lambda ' - \delta \lambda ', \theta_i ) ~ p( \theta_s )
~d \Omega ~d \lambda ' ~d \lambda . 
\end{eqnarray}

\medskip
 
\noindent
The parameters in the above equation are: the constant 
C = $B_{12} h {\lambda _i}^{-1} $, where
$B_{12}$ is the Einstein absorption coefficient, h is Planck's
constant, and $\lambda _i$ is the line center wavelength.
$N_1$(O$^{5+})$, is the ground state number density for O$^{5+}$. 
The O$^{5+}$ number density can be defined in terms of
the electron density, $N_e$, by using the following:
$N_1({\rm O^{5+}}) ~=~ 0.8 A_{\rm O} R_i(T_e ) N_e$,
where $A_{\rm O}$ is the total 
oxygen abundance relative to hydrogen; and $R_i(T_e )$ is the
ionization balance term for O$^{5+}$. The ionization balance
term is a function of the electron temperature (a true thermal temperature), $T_e$.  Using the
above substitutions in the equation for the combined emissivity, 
it can be shown that the abundance and ionization balance
terms cancel in taking the emissivity ratio $E_1/E_2$.
This is an advantage when determining the outflow velocity
from the emissivity ratio, since the uncertainties in these quantities
are not propagated to the uncertainties in the final velocity
determinations.

The third term, $\phi_r ( \lambda - \lambda _i )$, is the resonantly scattered coronal line profile which is a function of the small-scale ion velocity distribution in the direction of the incoming radiation, $f(w)$. Also specified are the absolute intensities of the incident disk radiation, $I_D$, as a function of wavelength, $\lambda '$, and the angle of incidence, $\theta _i$,
from the disk.  The Doppler shift $\delta \lambda '$ is a function of the
coronal outflow velocity $V_o$ (through the usual relation
$\delta \lambda ' = \lambda ' {V_o}/{c} $).  The phase
function parameter $p ( \theta _s )$ is the angular
dependence of the scattering process, where $\theta _s$ is the
angle between the incident and scattered radiation.  The
total resonance scattered emissivity involves an integration
over the solid angle, $\Omega$, which is the angle subtended by the solar disk.
The wavelength integrations are performed over the incident profile from the disk (primed) and the scattered (unprimed) profile in the corona.

%********************************************
In performing the computations for the coronal emission line profiles, 
it is more useful to work with velocity units instead of wavelength units.  In doing so,
the observed line profile is treated as a velocity distribution which has a
line width that is equal to the 1/e half-width of
the summed velocity distributions of all of the scatterers along
the line of sight.  Some of the line broadening in the direction along the line of sight is due to the projection of the bulk outflow along the line of sight.   This effect is included in the determination of the true line widths.  In a large polar coronal hole for example, the line profiles far from the plane of the sky will be red shifted if they are behind the plane of the sky and blue shifted if they are in the foreground. This additional effect on the profiles is taken into account with the {\it CORPRO} model.

We assume that the coronal velocity distributions are not isotropic but instead are bi-Maxwellian with 1/e velocity half widths, $w_\|$ and  $w_\bot$ \citep[e.g.,][]{cran99,akinari07}. The velocity distributions, w, have contributions from both thermal and non-thermal motions, such as turbulence or wave sloshing.  In general, the velocity distribution width ($w_\|$) in the direction parallel to the local magnetic field is not the same as the velocity distribution width ($w_\bot$) in the perpendicular direction \citep[see][]{kohl98}. We see from above that it is the  $w_\| $ component that provides the sensitivity for determining the coronal outflow velocity. 
Although $w_\| $ is not measured directly, it has been found to be highly constrained by several authors
\citep[e.g.,][]{cran99,xli98,FCK03}.  In particular for coronal holes, \citet{CPK08} found that by using an exhaustive search in parameter space that $T_\bot / T_\|  = 6 $ was the most probable anisotropy ratio for the oxygen kinetic temperatures, defined as $T_{\bot,\|} = m w_{\bot,\|} ^2 /(2k)$.  We use this value to determine the outflow velocities in coronal holes but show below (in section~\ref{uncert}) the effect on the outflow velocities if other values are used.  We use the same anisotropic temperature ratio for all bins in coronal hole regions, defined where the electron density is below $2 \times 10^{5}~cm^{-3}$.   For coronal streamers, where the higher densities imply that collisions should reduce the anisotropy, we use $T_\bot / T_\|  = 1 $ \citep[e.g., see][]{str02}.  We also assume that the temperature for all species (protons, electrons, and $O^{5+}$) are the same for streamers at our source surface height.   At heights above $2.3~R_{\sun}$ in streamers, \citet{FCK03} do provide evidence for temperature anisotropy, but only in mid-latitude streamers near solar minimum.

Before the outflow velocities can be produced, one needs to know the electron temperatures and the electron densities in the observed coronal regions in order to establish the baseline emissivities in equations (\ref{eqn_col}) and (\ref{eqn_res}).  We originally used an adjustable electron temperature for each region but found that this made a very small difference in the results so instead, we fixed this parameter to $T_e  =  1 \times 10^6 ~K$ for all regions. The electron densities for this work are derived from an  inversion of the LASCO C2 polarized brightness measurements of the white light corona \citep{van50}.  The specific implementation used the axisymmetric model of \citet{QuLam02} which is appropriate for this phase of the solar cycle (near solar minimum).  A more sophisticated implementation \citep[e.g.,][]{saez07} can be used for obtaining the coronal electron densities during more active phases of the solar cycle.  This is not warranted for the current work since the high spatial resolution of the LASCO-C2 images has been reduced to match the spatial bin size used for UVCS Carrington maps.  In addition, the original LASCO density maps for $2.5 R_{\sun}$ were adjusted to estimate the densities at the same height ($2.3 R_{\sun}$) that was used for the outflow velocity calculations. These densities were obtained by computing density ratios $N_{e}(2.3 R_{\sun})/N_{e}(2.5 R_{\sun})$ from previously studied streamers and coronal holes. The average value for this ratio was found to be 1.4 for streamers and 1.5 for coronal holes. \citet{susino08} suggest that LASCO derived densities in streamers may be may be too large compared to local densities derived using the \ion{O}{6} intensities.  While this may be true for the high latitude streamer that they examined, this correction is not applicable when looking through a horizontal streamer belt that exists during the solar minimum phase.

The reported outflow velocities are computed on a grid of 30 $\times$ 28 (latitude $\times$ longitude) bins using only west limb data. The outflow velocity in each bin is determined by creating a series of \ion{O}{6} profiles for a trial set of 26 outflow velocities that lie between 0 and $500~ km~s^{-1}$. The resonantly scattered and collisional emissivities are computed along the line of sight at each spatial bin, where we assume that the bulk of the emission comes from within $1R_{\sun}$ from the plane of the sky.  After the emissivities are summed to form the separate 103.2~nm and 103.6~nm intensities, we compute the intensity ratios that are compared to the observed \ion{O}{6} ratios. One advantage of using the line ratio is that the oxygen abundance drops out of the calculations. The most likely outflow velocity value for the spatial bin is the one that produces an \ion{O}{6} ratio which matches the observations.  All specified parameters are held fixed along the line of sight in order to reduce the number of iterations.  As is usual with forward modeling, the final outflow velocities may not be unique.

%%%%%%%%%%%%%%%%%%%%%%%%%%%%%%%%%%%%

\section{ELECTRON DENSITY AND OUTFLOW VELOCITY CARRINGTON MAPS} \label{nvmaps}

In the following sections, we present electron density maps made from white light polarized brightness data and, for the first time, outflow velocity maps obtained with the new Doppler dimming diagnostic code (see Section \ref{empmod}).
LASCO and UVCS daily synoptic images of the corona were used to make Carrington maps at selected heights.  Constructing the electron density maps is straightforward since the data reduction starts with fully two-dimensional images from the LASCO-C2 coronagraph \citep[e.g.,][]{bwsm99}. For each day, white light intensities or polarized brightness values are recorded, at a fixed height, for a complete $360 \arcdeg$ ~(in position angle) around the Sun.  A full Carrington map for a single rotation period can be made when data are placed in a rectangular grid with latitude and longitude bins.  To do this the observation times are converted to Carrington longitudes (assuming solid body rotation of the corona). The approximately 1-day interval between each observation corresponds to about $13.5 \arcdeg$ in longitude but we actually use an interval which depends on the variable speed of the SOHO spacecraft in its orbit around the Sun. 

Constructing the UVCS maps of spectroscopic parameters used to infer the outflow
velocities is not as straightforward as preparing the LASCO maps.  The main reason is that
UVCS uses a slit spectrometer with a narrow ($\sim 1\arcmin \times 40 \arcmin$) field of
view of the corona.  An additional step requires the construction of two-dimensional plane-of-the-sky
"images" of the corona for the UVCS observations.  To build up a complete coronal image,
\ion{O}{6} and H~I Ly$\alpha$ profile measurements are obtained with the UVCS field of view positioned at several different heights between 1.5 and $3.5~R_{\sun}$ while the instrument remains at a fixed roll angle (i.e., position angle).  A complete raster of the full corona (called a `synoptic image') is produced by making similar height scans at a total of eight roll angles about the Sun, with $45 \arcdeg$ intervals between each roll.  Fits are made to the \ion{O}{6} profiles in each UVCS spatial bin so that intensity and line width information can be obtained for each point in the sky. The UVCS data (intensities and line widths) are then interpolated along a pole-to-pole arc at a fixed distance from Sun-center in order to make a uniform data set with equal latitude intervals.  It should be noted that the coronal line widths are corrected for the instrument profile function and the spectrometer silt width.  At this stage, the data from successive daily image maps are time tagged and converted to Carrington longitude just as for the density Carrington maps.  More details about the preparation of the UVCS Carrington maps are described in \citet{str97} and \citet{pan99}.

Once the Carrington maps of the UV profile data (intensities and 1/e line widths only)
and the electron densities are computed, there is enough information to estimate the bulk
outflow velocities for each latitude/longitude bin of the Carrington maps.  The outflow
velocities are computed using the Doppler dimming model described in section \ref{empmod}.
The only other information needed is the known atomic parameters for the collisional and radiative excitation rates \citep{GW70, noc87}; the ionization equilibrium values for the atomic levels \citep{mazz98}; and the intensities and line widths for the disk spectral lines \citep{noc87,xli98}.

%%%%%%%%%%%%%%%%%%%%%%%%%%%%%%%%%%%%%
\subsection{Carrington Maps for CR 1912 and 1932} \label{smaps}

\begin{figure}
\plottwo{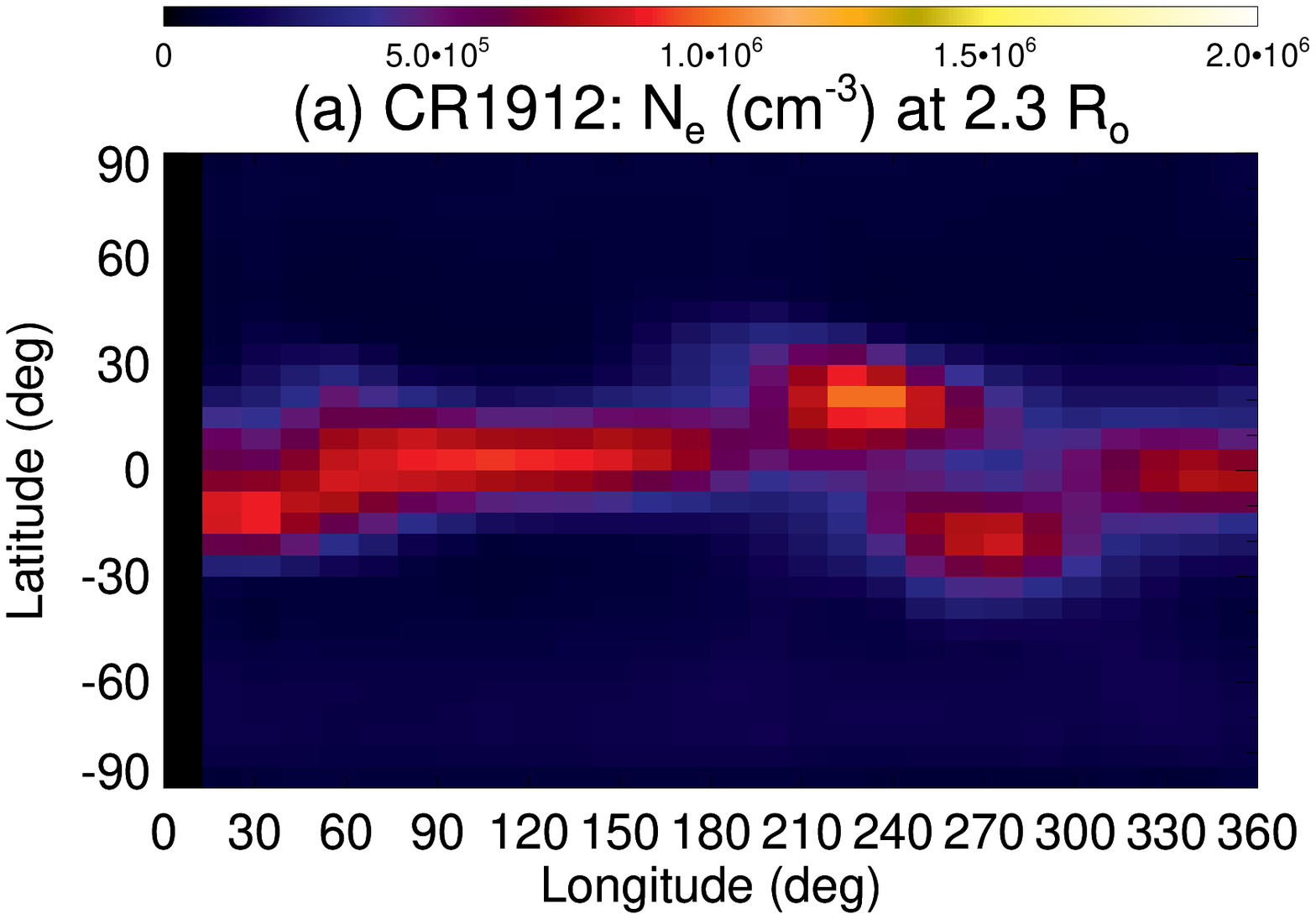}{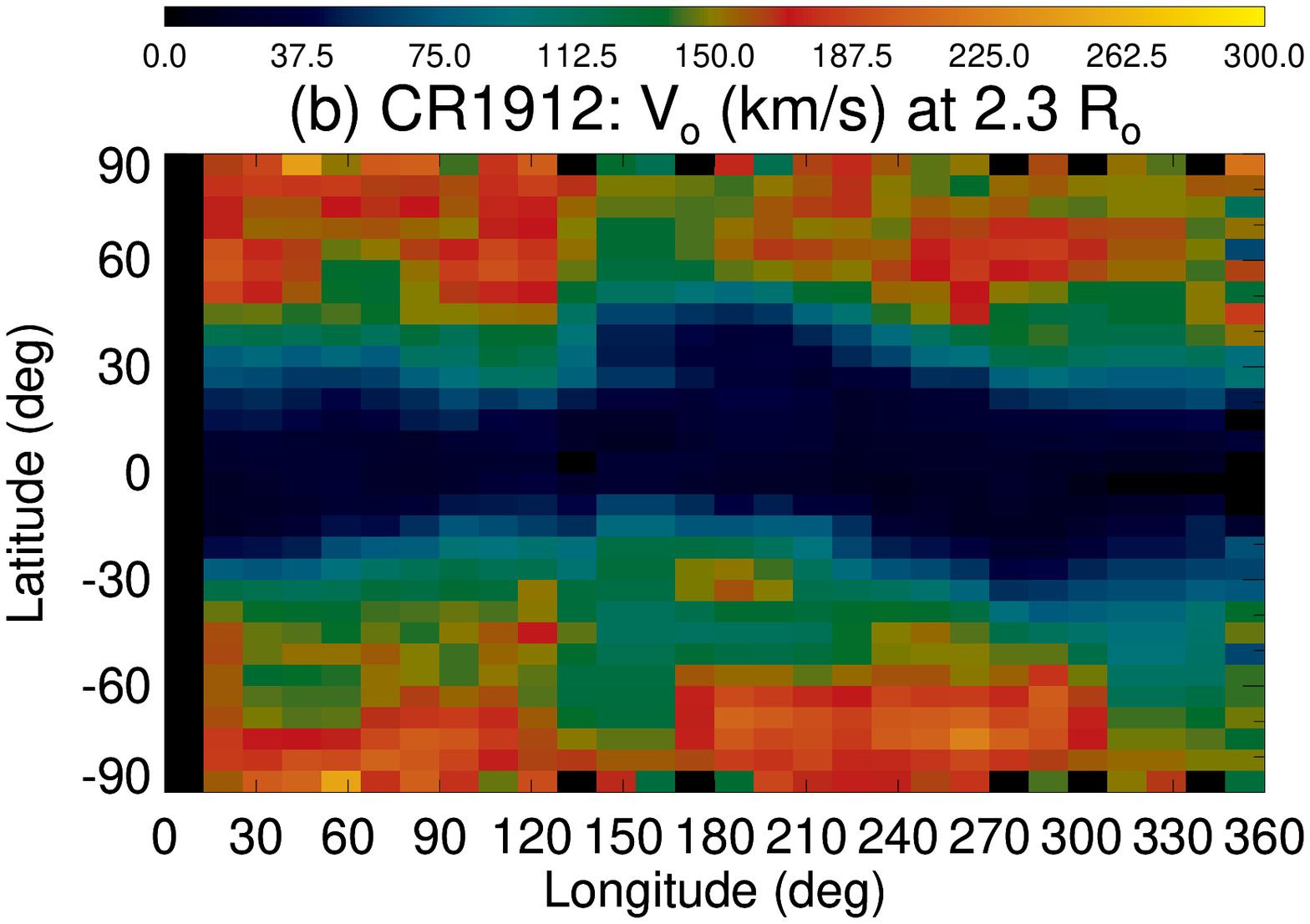}
\plottwo{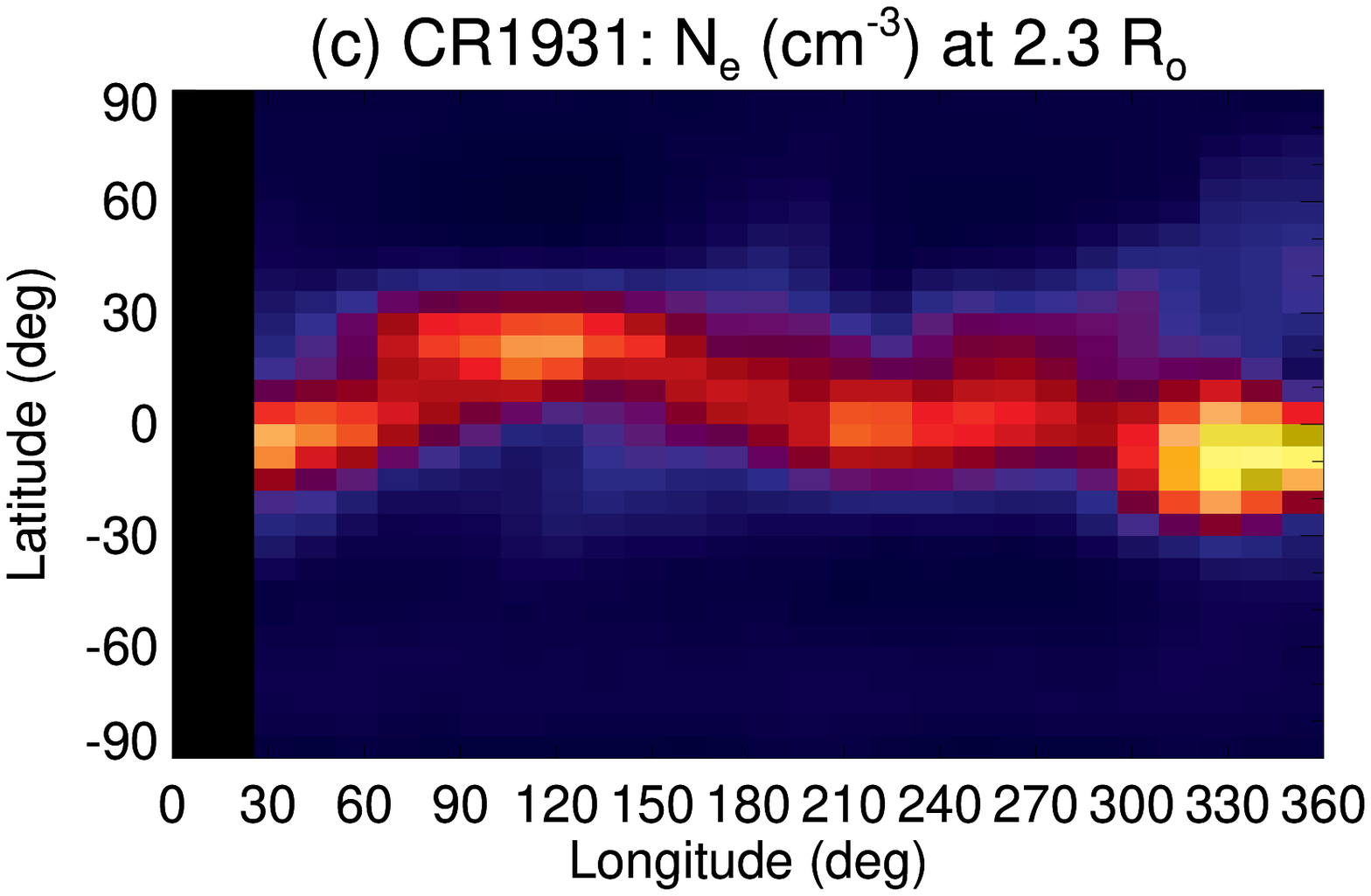}{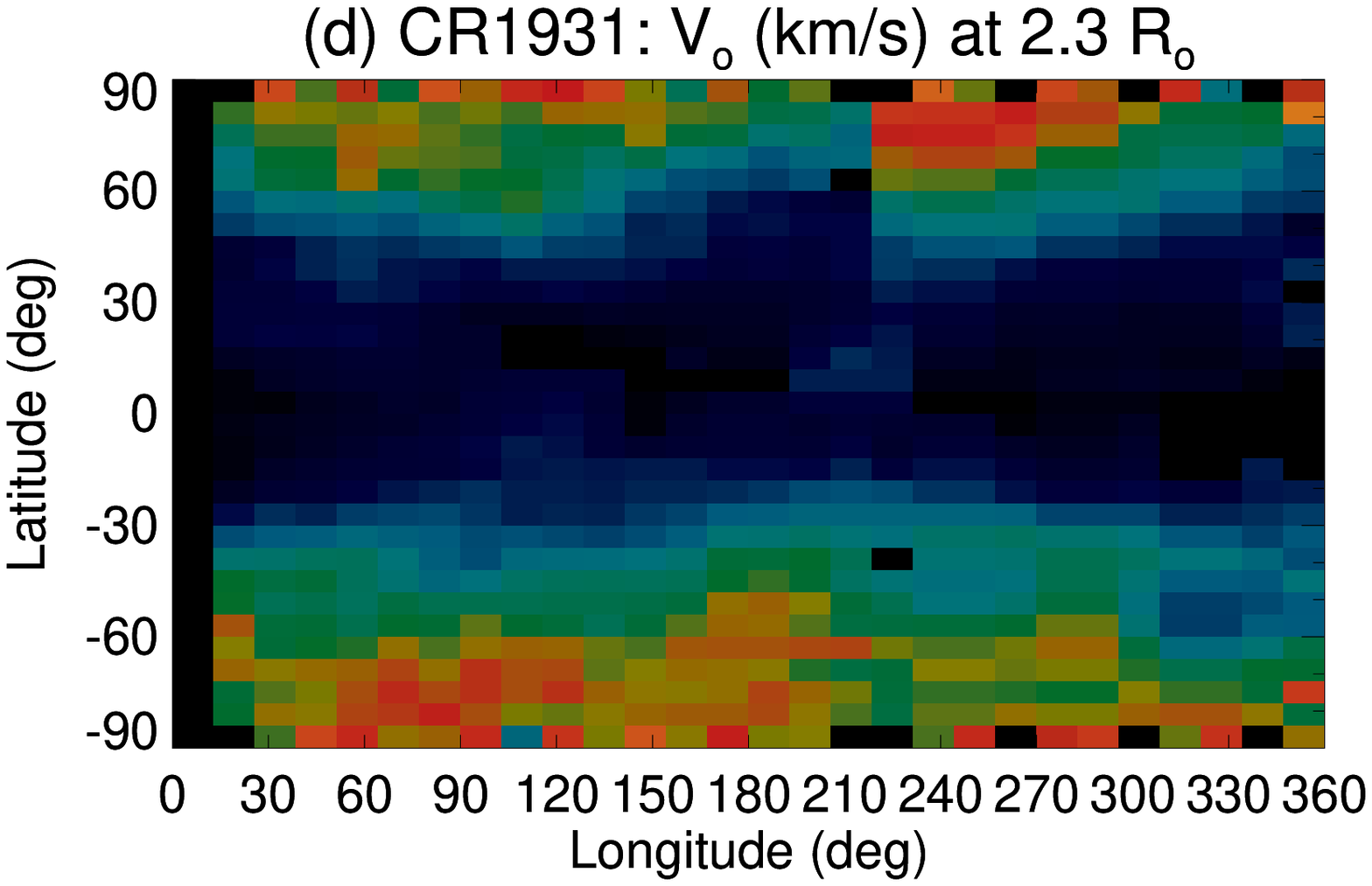}
\caption{Maps of electron density and solar wind outflow speeds shown for a heliocentric height of 2.3~R$_{\sun}$.  Panels (a) and (b) are for CR 1912 (Aug 1996) and panels (c) and (d) are for CR 1931 (Jan 1998). \label{fig_ne_v2x2}}
\end{figure}

Figure~\ref{fig_ne_v2x2} shows two representative maps\footnote{Individual density and outflow velocity maps for Carrington Rotations 1909--1937 can be found at $\rm{http://www.cfa.harvard.edu/\sim strachan/ModVal/}$} for the electron densities and O$^{5+}$ outflow speeds at 2.3~R$_{\sun}$.  Panels (a) and (b) are for Carrington Rotation 1912 (near solar minimum at the start of cycle 23) while Panels (c) and (d) are for CR 1931 (17 months later) when solar activity has increased and the coronal magnetic current sheet has become more warped.  The color bar above the density plots indicates that the coronal hole regions have number densities less than $5 \times 10^{5} ~cm^{-3}$ at the selected height.  Coronal streamers which form the bright orange/red band in both panels (a) and (c) can have densities as high as $2 \times 10^{6} ~cm^{-3}$. All of the densities are shown for 2.3 R$_{\sun}$ and were computed by scaling the density values from the LASCO density maps at 2.5 R$_{\sun}$ as previously described above.

\placefigure{fig_ne_v2x2}

In these figures, the LASCO Carrington maps have been resampled to match the resolution of the UVCS Carrington maps, which use a binning of 30 $\times$ 28 (latitude $\times$ longitude). Note that the apparent break in the streamer belt near $270 \degr$ longitude in Figure~\ref{fig_ne_v2x2}(a) is caused by the deflection of the heliospheric current sheet by an active region on the disk.   The apparent tilt of the diagonal streamer arms at mid latitudes, e.g., in Figure~\ref{fig_ne_v2x2}(a) between $180 \degr$ and $270 \degr$, has been shown by \citet{Wang1997} to be related to the tilt angle of the solar rotation axis. The direction of the tilt depends on whether the solar rotation axis is in front of or behind the plane of the Sun and on whether the map is constructed from East-limb or West-limb data.  While the tilts for CR 1931 are faint in the density plots (Figure~\ref{fig_ne_v2x2}[c]), they are very pronounced in the velocity maps (Figure~\ref{fig_ne_v2x2}[d]). 

%
%%%%%%%%%%%%  Start of section on Vout %%%%%%%%%%%
%
The outflow velocities for Carrington Rotations 1912 and 1931 are shown in Figure~\ref{fig_ne_v2x2}(b) and (d), respectively. The velocity maps have been interpolated to fill in for some days where there were no observations from either UVCS or LASCO.  The missing days (columns) are identifiable on the maps by the columns with black bins at both the top and bottom rows (at $90 \arcdeg$ and $-90 \arcdeg$).  Interpolations for missing data are linearly applied in both the longitudinal (temporal) and latitudinal directions.  Note that the dark, low-velocity regions in the outflow velocity maps correspond to the light colored, high-density regions in the density maps.  This shows that, already by 2.3~R$_{\sun}$, the highest speed outflows come from the broad coronal holes surrounding both poles.  An interesting result is that the latitudinal width of the slow speed wind is considerably wider at CR 1931 than it is at CR 1912 (the heart of the solar minimum); however the thickness of the streamer belt (in the density) for both time periods is comparable.  We will return to this point in section~\ref{allmaps} below.

%===========================================
\subsection{Contiguous Carrington Maps for Cycle 23 minimum} \label{allmaps}

\begin{figure}
\plottwo{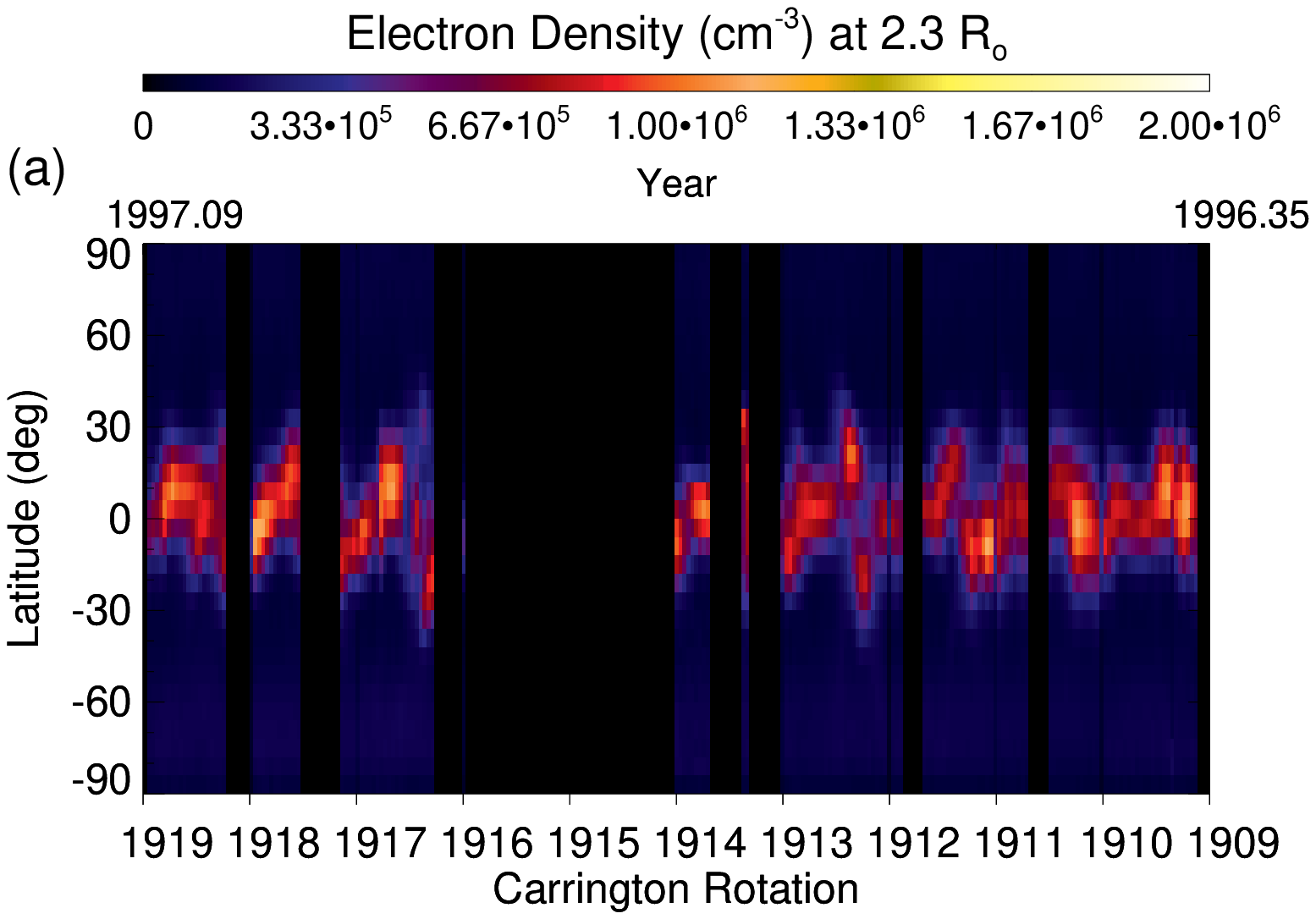}{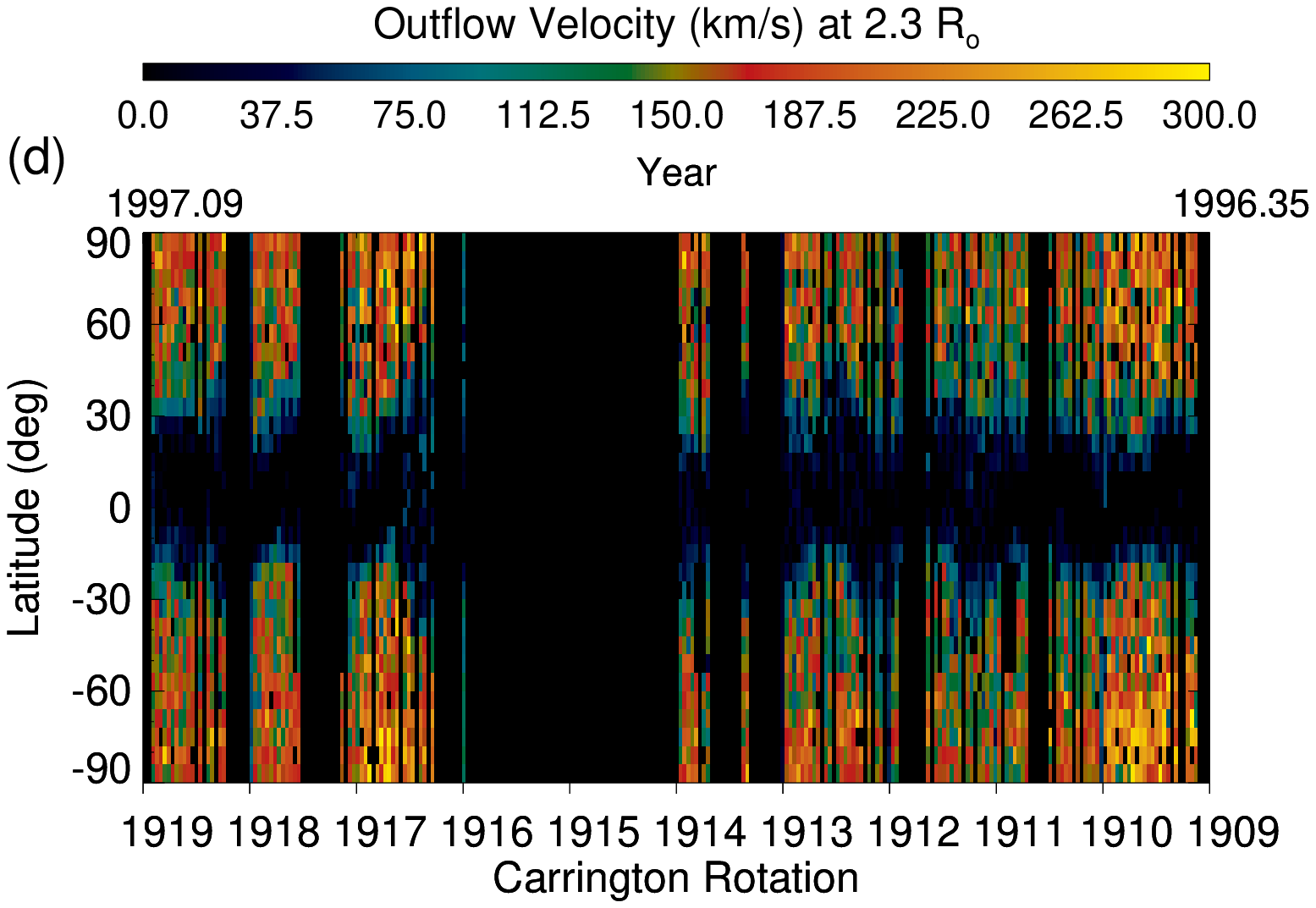}
\plottwo{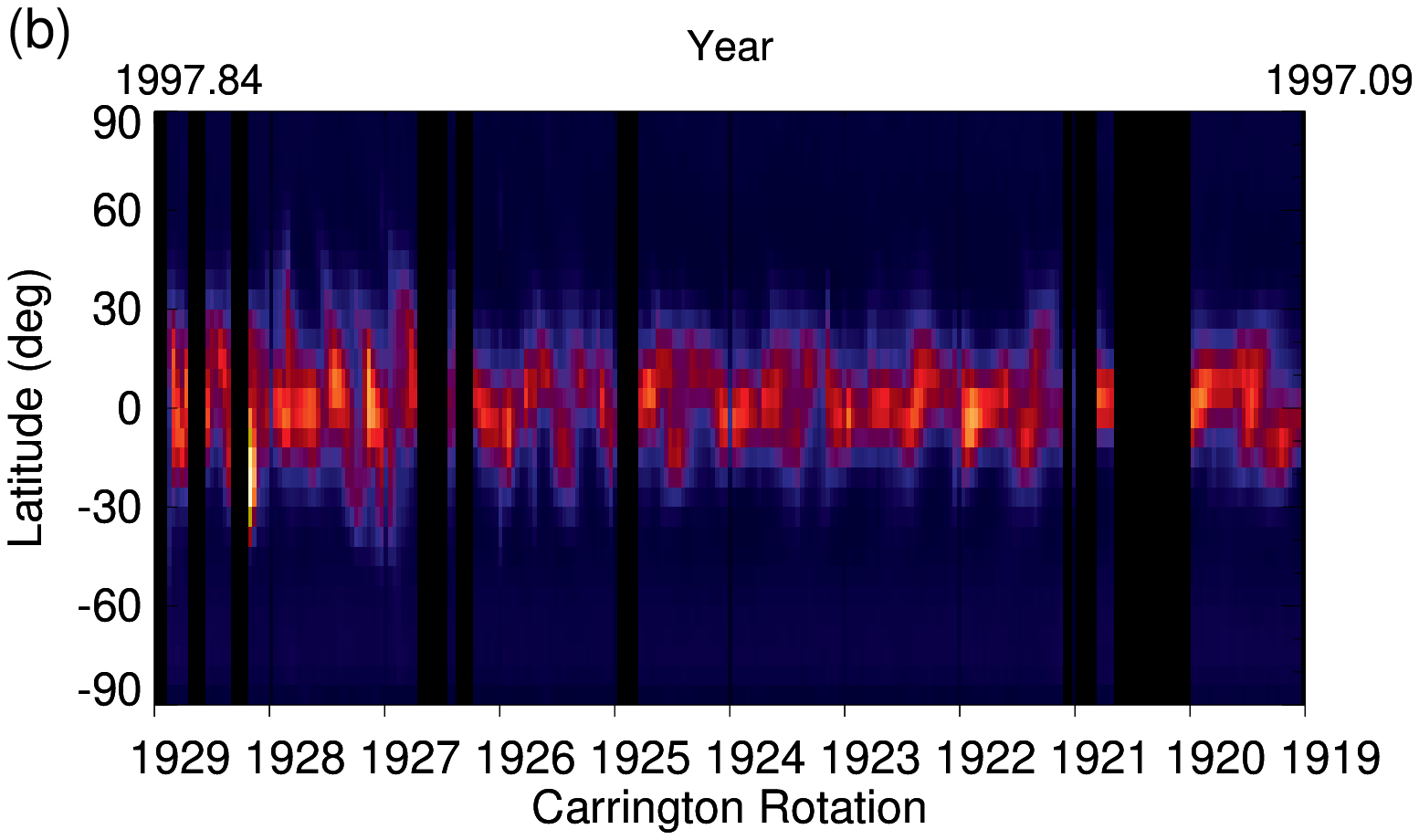}{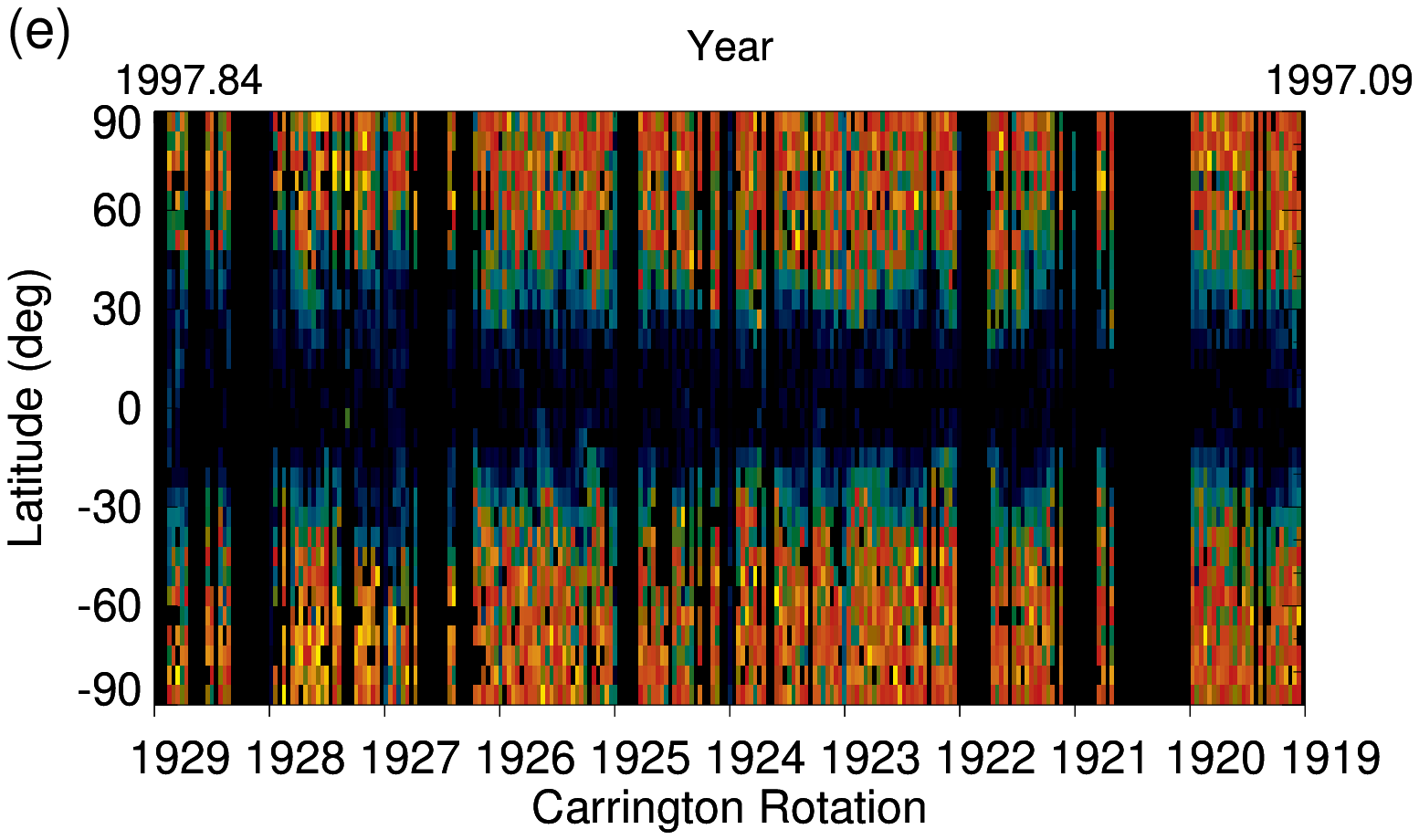}
\plottwo{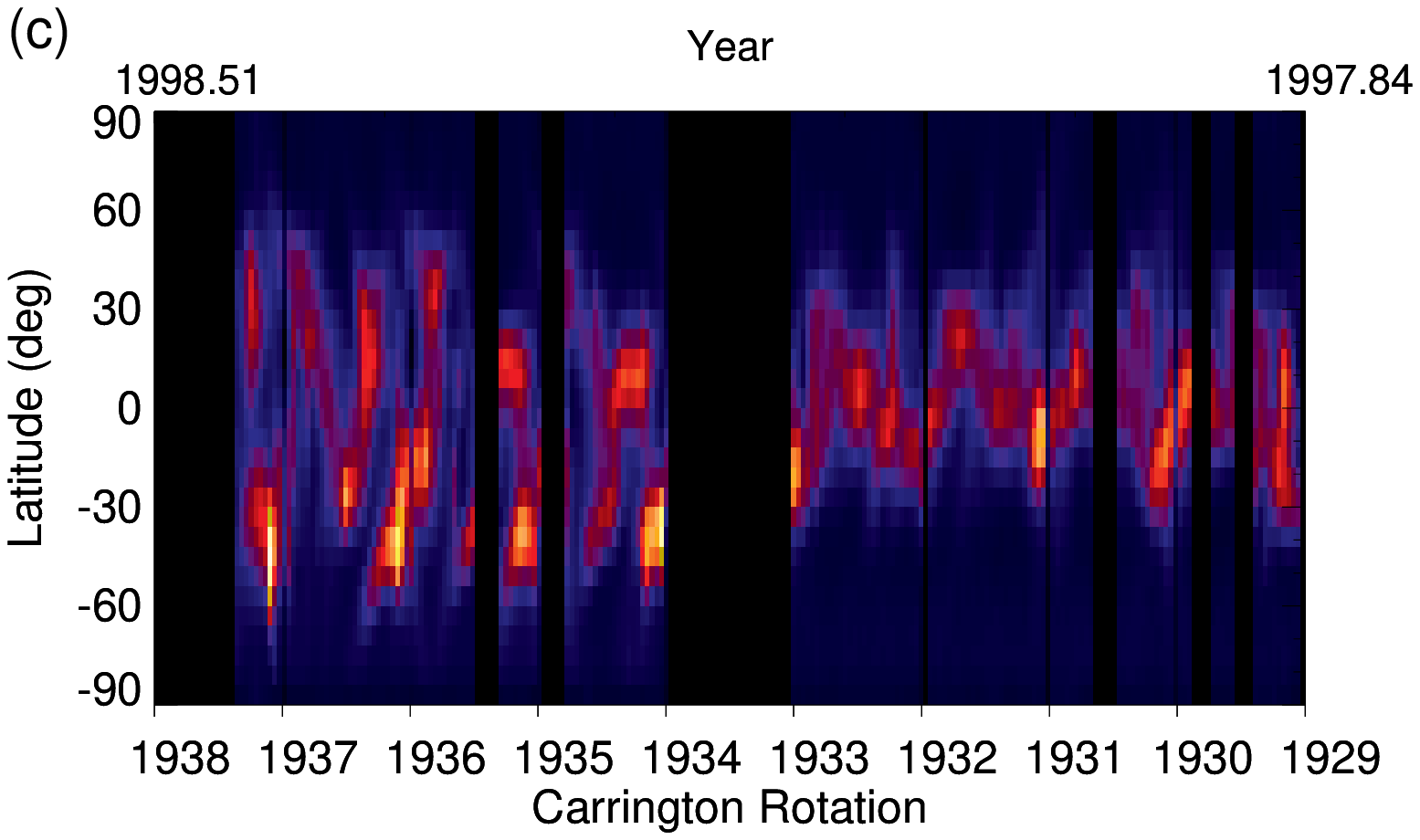}{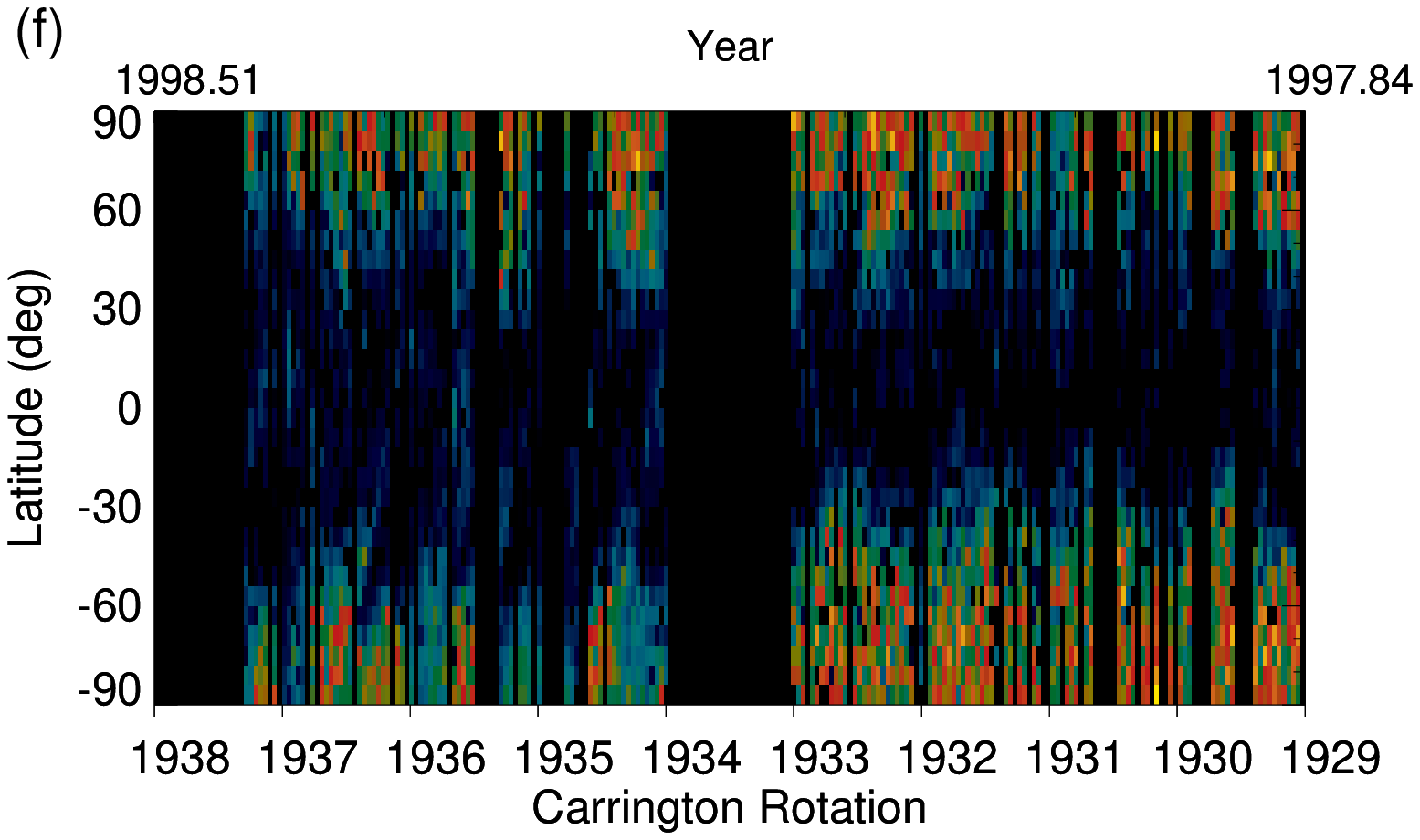}
\caption{Changes in the electron density (left panels) and outflow speed (right panels) as a function of latitude and time for Carrington Rotations 1909 to 1937.  The two different color bars above the top plots are used to quantify the density and velocity color scales for the respective column of panels.   A  color version of this figure can be found in the online electronic version of the paper.
\label{fig_vel_all}}
\end{figure}

In order to provide a global perspective of the period near the minimum at the start of cycle 23, we place all of the available Carrington maps for electron densities and outflow velocities side-by-side.   The maps for 2.3 R$_{\sun}$ are shown in Figures~\ref{fig_vel_all} for the electron densities (left side) and for the coronal outflow velocities (right side).  The entire period is divided into three parts: CR~1909 -- 1918 (top panels), CR~1919 -- 1928 (middle panels) and CR~1929 -- 1937 (bottom panels). To preserve the standard longitude orientation for each rotation period (longitude increases from left to right on the horizontal axis), the maps were constructed so that observation time increases to the left.  The discontinuities in the maps are due to extended periods ($>$ 2 days) without valid data.  The longest time gaps (e.g., CR~1915 -- 1916) are due to periods when the SOHO spacecraft temporarily lost its Sun-pointing attitude control and therefore, no observations were made.  Other shorter gaps ($\sim 2$ days in duration) are due to several factors.  Some of these are for periods when the UVCS synoptic observations were not made.  Other gaps, especially those that extend in the latitude direction, are due to instances where the data were missing or corrupted.

\placefigure{fig_vel_all}

Although the vertical height of each map in Figure~\ref{fig_vel_all} is the same as that for Figure~\ref{fig_ne_v2x2}, the horizontal axis has been compressed considerably. The maps reveal the evolution of the density and velocity structures in the transition from solar minimum to the rising phase of the solar cycle.  The figures clearly show a gradual drift towards the poles of the equatorial high density, slow speed region as is expected during the progression away from solar minimum.  This latitudinal spread of the low velocity streamer belt toward the poles is not uniform and there is an indication that the minimum width of the slow speed belt did not occur during sunspot minimum in Carrington Rotation 1909 in May 1996 \citep{SIDC10} but possibly as late as CR~1922 (shown in the middle panels). The latitudinal width of the low speed belt appears nearly constant from approximately CR~1921 to CR~1926, indicating a broad minimum in the belt thickness.  This broad minimum is most easily seen in the middle density map shown in Figure~\ref{fig_vel_all}.  

Before describing how the corona evolves in time, we present the mean properties (as a function of latitude) of several solar wind parameters in the plots on the left side of Figure \ref{fig_lat_var_all}.  These data are useful for providing a quantitative characterization for the mean conditions during the solar minimum period at the start of cycle 23.  Figure \ref{fig_lat_var_all} shows the following:  a) the solar wind outflow velocity, b) electron density, and c) the particle flux (all averaged over Carrington Rotation periods 1909-1925) for each latitude bin.  The statistical $1~\sigma$ variations for the data points in each plot are shown as vertical lines and are generally small, i.e., about the size of the symbols for each plot. The error bars give an indication of the variation of the data but not the overall uncertainties in the parameters which are discussed in section \ref{uncert}.

\placefigure{fig_lat_var_all}
 
At the heart of solar minimum (CR~1909 -- 1925), we see a nearly symmetrical corona in both
the outflow velocity and electron densities as a function of latitude.  The plots are not
inverses of each other; there are some subtle differences between them.  First of all, it
should be noticed that the outflow velocity minimum is flat compared to the sharper peak
in the density data.  This is somewhat artificial since the averages were calculated by
using only the {\it nonzero} velocity values in the streamer belt (we are mainly
interested in the open field regions with solar wind outflow). Another noticeable 
difference is that there is a steeper outflow velocity slope in the transition between the
polar coronal holes and the low latitude streamer belt, when compared to the density
transition.  The implication is that, at least at the selected source surface height, both polar coronal holes have a gradient in their velocity profile.  The smoother density transition could also be attributed to the line of sight integration.

It is interesting to compare the velocity vs. latitude plot in Figure \ref{fig_lat_var_all}(a) with the Ulysses proton velocity vs. latitude plot in \citet{McComas00}.  The Ulysses plot has a much larger range of latitudes with fast wind since the coronal holes are still over expanding above $2.3~R_{\sun}$.  The velocity gradient for the fast solar wind ($V > 700~km~s^{-1}$) is $0.95~km~s^{-1} deg^{-1}$ above $36 \arcdeg$ latitude.  Our data has a mean velocity gradient (using data from both poles) of $0.68 ~km~s^{-1} deg^{-1}$   for latitudes above $54 \arcdeg$ , which is outside of the slow wind belt.  The gradient in the south pole alone is $0.85  ~km~s^{-1} deg^{-1}$  (negative). This is larger than the velocity gradient for the north pole using our data and is closer to the Ulysses value.  

Also notice in Figure \ref{fig_lat_var_all}(c) that the mean particle flux for the escaping solar wind is approximately $1.8 \times 10^{12} ~cm^{-2} ~s^{-1}$ for all latitudes which include the streamer belt and the polar coronal holes.  The bump in the particle flux between $-90 \arcdeg$ and $-30 \arcdeg$ is probably real.  There appears to be a smaller one  in the northern hemisphere as well.  We will comment more about these features in section~\ref{discuss}, when we discuss how the plasma parameters for one rotation (CR~1909) compares to a theoretical model.

%%%%%%%%%%%%%%%%%%%%%%%%%%%%%%%%%%%%%%
\subsection{Evolution of the Source Surface Parameters at Different Latitudes} \label{ss_evol}

In this section we describe the spatial and temporal variation of the coronal data with the emphasis mainly on the outflow velocities since these are the newer results.  The corona had a simple structure during the cycle 23 minimum with a nearly continuous streamer belt and two large polar coronal holes that were present throughout the period of this study.  In order to better observe the long term trends, we use averages for each Carrington rotation at three selected latitudes:  $90 \arcdeg$, $60 \arcdeg$, and $30 \arcdeg$.  The north and south hemisphere data are similar except for higher densities at the south pole (see Figure~\ref{fig_lat_var_all} above).  In Figure~\ref{fig_v_3lat_crn} we present plots for the mean values of the outflow velocity (V), plasma density (N), and proton kinetic temperature ($T_p$) in the three stated latitude bands.  The data for $0 \arcdeg$ are not plotted since most of the values obtained for the outflow velocity were too small to measure.  The large number of unmeasurable small outflow velocities makes the concept of an average value not very useful at the equator.

\placefigure{fig_v_3lat_crn}

The rotation-averaged outflow velocities at high latitudes ($60 \arcdeg$ and at $90
\arcdeg$) are fairly constant up to CR~1925, after which they start to decline with time. At  $30
\arcdeg$ latitude, the decline in the outflow velocity starts earlier at about CR~1920.  The
decrease in the solar wind outflow velocity may be related to the appearance of dense
streamers at low latitudes (see the density plot at $30 \arcdeg$ in the middle panel).
Contrary to the density rise for the $30 \arcdeg$ plot, the high latitude plots of density
vs. time are nearly flat.  The nearly constant density with the corresponding increasing
proton temperatures (bottom panel) for the high latitude plots may partially explain the
decrease in outflow velocity over the poles.  If one assumes a time steady average energy
input into the coronal holes, it appears that over time, more energy is converted into
heating the coronal plasma rather that increasing the outflow velocity.  On the other
hand, the bottom panel in Figure~\ref{fig_v_3lat_crn} shows that the proton kinetic
temperatures in the high density streamer regions {\it decrease} with time, especially
after CR~1925.  The lower kinetic temperatures could be explained by the fact that the increase in density of the low latitude regions leads to an increase in the radiative cooling of the streamer plasmas.   Another  effect that could be important is that the denser streamer plasmas have a smaller non-thermal temperature component (compared to coronal holes) and this reduces the overall proton kinetic temperature in these regions.  This reduction of the non-thermal component in streamers is qualitatively consistent with a one-fluid model which shows that the non-thermal perpendicular velocities (produced by Alfven waves in this case) are lower in the streamer legs compared to the values in a coronal hole \citep[see Figure 11 in][]{cran07}.

%%%%%%%%%%%%%%%%%%%%%%%%%%%%%%%%%%%%%%
\subsection{Solar Wind Source Regions} \label{ch_bound}

In order to better understand changes of the intrinsic properties of the solar wind source regions, we need a method to identify coronal hole regions that are uncontaminated by streamer material that may be along the same line of sight.  This can be accomplished by using an outflow velocity criterion or a density criterion to distinguish coronal hole regions from streamers.  We show below that both type of criteria give similar results for defining the pure coronal hole regions.  For the first method , we examine the distribution of outflow velocities from all spatial bins on the source surface maps for the entire period from CR~1909--1937.  Figure~\ref{fig_histo_vel} shows a histogram plot of the outflow velocities. The distribution is clearly double peaked with a low velocity peak centered at $\sim 30 ~km~s^{-1}$ and a high velocity peak centered at $\sim 150 ~km~s^{-1}$.   For convenience, we assume that the
two velocity groups represent the fast and slow wind components at the  $2.3~R_{\sun}$ source surface. It should be noted that the number of counts in the lowest outflow velocity bin  do not include pixels with V = $0 ~km~s^{-1}$.

\placefigure{fig_histo_vel}

The overlap between the two distributions occurs at roughly $\sim 90 ~km~s^{-1}$ and so we
will use this as the criterion for separating the fast and slow speed source regions.  Since we know that the fastest solar wind originates in polar holes, we can use the velocity break point as the separation between coronal hole and non-coronal hole source regions.  We can compute the mean latitude bin for the  transition from fast to slow wind by starting at the pole deep inside of the coronal hole and then move to progressively lower latitude bins until we reach a bin where the outflow velocity falls below $90 ~km~s^{-1}$.  The mean latitude of this boundary for each Carrington Rotation is shown as the filled circles connected by the solid lines in Figure~\ref{fig_nch_lat_crn}.  The uncertainty of the latitude location of the boundary is computed by combining the $1 \sigma$ uncertainty due to the latitude binning of the data ($\pm 3 \arcdeg$) with the uncertainty in the velocity at the cutoff location. 

\placefigure{fig_nch_lat_crn}

For comparison, the open square symbols show the coronal hole boundary (CHB) using the method of \citet{gut96} where the boundary is found by plotting the electron density profile from pole to pole (along a constant longitude).  The latitudinal profile for the density is approximately Gaussian with the peak density near the equator in the streamer belt and the base of the profile in the polar coronal holes (see for example, the electron density profile in  Figure~\ref{fig_lat_var_all}b).  The coronal hole densities form a baseline above zero which is approximately 15\% of the peak density in the streamer belt \citep{gut96}.  It is remarkable that the coronal hole boundary determined with this density threshold is nearly identical to the boundary determined by using the outflow velocities.  The fact that the results from the two techniques are in close agreement provides some confidence in the identification of the coronal hole boundaries for our Carrington maps. 

Once the coronal hole boundary is defined at $2.3~ R_{\sun}$, in this case for the northern hemisphere, we can examine how the size of the coronal hole changes over time.  Figure~\ref{fig_nch_lat_crn} shows that the CHB was fairly constant at $\sim30 \arcdeg$ latitude until Carrington Rotation 1925; after this the boundary increases rapidly in latitude to about $60 \arcdeg$ by CR~1937.  Therefore, the best period for characterizing the steady-state plasma properties of the coronal hole for the case of a relatively stable geometric configuration is during the period CR~1909 --1925.  After CR~1925, the latitude of the CHB increases rapidly and so the intrinsic changes of the coronal hole plasma properties would be more difficult to interpret since they could be masked by the fact that the volume of the coronal hole is also changing.

Also plotted in Figure~\ref{fig_nch_lat_crn} is the mean
latitude of the coronal hole boundary on the disk (at $r \approx 1~R_{\sun}$).  The boundary is indicated by open circles using data obtained from
He~I 1083.0 nm observations \citep{har02}.  The uncertainties for these data are not shown since they are smaller than the symbol size.  The authors estimate that the boundaries are determined to within $1 \arcdeg$ for most of the data and to within $3 \arcdeg -5 \arcdeg$ for times when the boundary is less clear.  As an average uncertainty, we use $2 \arcdeg$ for these data. There is a clear indication that the polar
coronal hole has a super-radial expansion since the CHB on the disk is nearly constant at
$\sim 60 \arcdeg$ latitude while the same CHB at $2.3~R_{\sun}$ (solid circles) is at much
lower latitudes.  However, this is not true for the entire period.  At the time of CR
1937 the coronal hole expansion from the disk to the source surface becomes essentially radial since the CHB at both heights are at approximately the same latitude.  

The surface area of the coronal hole on the disk and at $2.3~R_{\sun}$, both as a function of time, are shown explicitly in Figure~\ref{fig_ch_area_fexp}.  
The plot in the left panel shows the rapid decline in the coronal hole surface area (in square solar radii) after the time period near CR 1925.  A factor of 10 multiplier is used to plot the area of the coronal hole on the disk in order to show the details of its variation with time.  The error bars indicate the estimated $1 \sigma$  uncertainty in the calculation for the areas. These are determined by combining, in quadrature, the uncertainties in the observation height
%($\sim 0.05~ R_{\sun}$)
with the uncertainties for the CHB latitude.  Again, the error bars for the disk data are omitted as they are smaller than the plotted symbols.

 \citet{MJ77} define the areal expansion factor $f_{A}$ at the source surface height, $r_{ss}$, as
\begin{equation} \label{eqn_fAexp}
f_{A}(r_{ss}) ~=~ \frac{A_{CH}(r_{ss})}{A_{CH}(r_o)} \frac{{r_o}^2}{{r_{ss}}^2}  ,
\end{equation}
where $A_{CH}(r_{ss})$ is the coronal hole area on a sphere of radius $r_{ss}$ and
 $A_{CH}(r_o)$ is the coronal hole area at the base height,  $r_{o} \approx 1R_{\sun}$. (Note that we take $r_{o}$ to be above the canopy structures at the coronal base where there is an initial rapid expansion of the flux tubes rising from the photosphere.)
The right panel of Figure~\ref{fig_ch_area_fexp} shows the calculated values of $f_{A}$ for the north coronal hole using the data for $A_{CH}(r_{ss})$ and $A_{CH}(r_o)$ shown in the left panel.

\placefigure{fig_ch_area_fexp}

An interesting feature of the areal expansion factor plot,  is that $f_{A}$ decreases with an approximately linear relationship with time, starting midway through the data period.  It reaches a value of
$\sim 1$ at the end of the period under study.  The data suggest that the
areal expansion factors of polar coronal holes are not constant and there is at least one
period in the solar cycle where $f_{A} = 1$, i.e., the expansion with height goes as
$r^2$.  However, unlike the suggestion of \citet{woo99}, our data show that the areal expansion of the polar coronal hole is super-radial during the deepest part of solar minimum, when the coronal holes are the largest.

A summary of the changes with time of the plasma parameters and mean size of the northern polar coronal hole is presented in Table~\ref{tab_nch_params}.  Mean and standard deviations for the outflow velocity, electron density, and proton kinetic temperature are shown for selected latitudes and Carrington rotations.  (The standard deviations are deviations from the mean values calculated by averaging over all longitude bins for the fixed latitude and specified Carrington rotation.)  As previously mentioned, there is little variation in the mean values of the plasma parameters between CR~1909 and CR~1925 (see Figure~\ref{fig_v_3lat_crn}). Consequently, we have omitted any intermediate Carrington rotations between these two endpoints.  We suggest that the changes that occur after CR~1925 can be characterized as the start of the rising phase of the solar cycle.  The data for the plasma parameters are tabulated for $30^{\circ}$, $60^{\circ}$, and $90^{\circ}$ latitudes to show their latitudinal dependences.  Mean parameters for the southern latitudes are similar except at the south pole which appears to be contaminated by foreground/background high density structures (possibly plumes or streamers). The geometric parameters that describe the mean latitude of the  coronal hole boundary, the coronal hole size, and its areal expansion factor (from the disk to the source surface) are shown in the last three rows of the table.   These parameters are shown for  $2.3~R_{\odot}$ in the last three lines of Table~\ref{tab_nch_params}.  Although the $\pm$ uncertainties for each quantity are sometimes unequal, only the average of the positive and negative uncertainties are shown in parentheses next to the mean quantities.          

\placetable{tab_nch_params}

%%%%%%%%%%%%%%%%%%%%%%%%%%%%%%%%%%
\subsection{Uncertainties of the Plasma Parameters} \label{uncert}

The standard deviations shown in Figures \ref{fig_lat_var_all}, \ref{fig_v_3lat_crn}, and Table \ref{tab_nch_params} provide an indication of the variation in the data from day to day in one Carrington rotation or the variation of the mean values over several rotations.  Estimates for the uncertainties in the each of the individual plasma parameters are described in this section. 

We estimate the electron density uncertainties to be about $\pm 30\%$. 
This range should cover the possibility of lower electron densities that are
derived from using the \ion{O}{6} intensities instead of the white light data
\citep{susino08,Abbo10}. 
The uncertainties for the proton perpendicular velocity distribution ($w_{\bot}$), which depends on the H Ly$\alpha$ line widths, are $< 2\%$.  The uncertainties in $w_{\bot}$, for $O^{5+}$ are about 20\% based on the uncertainties of the \ion{O}{6} line widths determined from UVCS profile observations.  Two more parameters in the coronal model are less well known:  the electron temperature, $T_e$ and the  O$^{5+}$ parallel velocity distribution width, $w_\|$.  As mentioned above, there is only a weak dependence on $T_e$ when using the \ion{O}{6} ratios and so it can be ignored. The uncertainty in $w_\|$ (which is related to the ion kinetic temperature,  $T_\|$) and its effect on the outflow velocity results are described below.

The uncertainty in the outflow velocity, V,  depends on the uncertainties in both the modeled \ion{O}{6} intensity ratios, which depend on the plasma parameters mentioned above, and the observed intensity ratios.  We can estimate the uncertainty in V by running the CORPRO model with the upper and lower $1 \sigma$ uncertainties in the input plasma parameters.  These model results can be compared to the observed ratios with their uncertainties added.  The observed intensity ratio depends mainly on the counting statistics of the individual \ion{O}{6} line intensities. The radiometric calibration factor \citep[with an uncertainty of 15\%][]{LDG96,LDG02} drops out when computing the intensity ratio.  Using the background corrected counts of the UVCS synoptic observing program, we get typical values for the $1 \sigma$ uncertainty in the observed \ion{O}{6} intensity ratio of 7\% in coronal streamers and 11\% in coronal holes at $2.3~R_{\sun}$. For intermediate, quiet sun regions we use 9\%.  The statistical uncertainty in the observed \ion{O}{6} intensity ratio has the second largest effect on the uncertainty for the outflow velocity.

The largest effect on the outflow velocity result is the uncertainty in $w_\|$ for the $O^{5+}$ ions, which can be parameterized with $T_\|$. Because of its impact on the outflow velocities in coronal holes, we treat this parameter separately.  In Table~\ref{tab_v_uncert} we show the estimated $\pm 1 \sigma$ uncertainties in the outflow velocity in three different velocity regimes. The uncertainties in the outflow velocities are determined by combining in quadrature the resulting changes in V due the $\pm 1 \sigma$ uncertainties in $N_e$, $T_{\bot}$, and the observed \ion{O}{6} ratio.  Because of the large sensitivity to $T_{\|}$, we show the results for three different values for $T_{\|}$ in coronal holes.  (We use $T_{\|} = T_{\bot}$ which is fixed for all cases in the streamer belt.) The baseline model in Table~\ref{tab_v_uncert} is the one where $T_{\|} ~=~ T_{\bot}/6$. We use this for all of the results in this paper.  The models where $T_{\|} ~=~ T_{\bot}/10$ and $T_{\bot}$ are thought to be very unlikely in the actual corona based on an exhaustive parameter study of a polar coronal hole by \citet{CPK08}. For the streamer belt, we use $T_{\|} ~=~ T_{\bot}$, based on the higher collision rate expected in the denser streamer plasma.

Occasionally, there is more than one velocity solution which gives the same modeled \ion{O}{6} intensity ratio. This can occur in a coronal hole when there is pumping of the \ion{O}{6} 103.7 nm line by the nearby \ion{C}{2} lines.  In these cases, we start with the outflow velocity determined in streamer belt along the same longitude and add the constraint that the outflow velocity varies smoothly when going from the streamer to the coronal hole.   We assume that the streamer belt and adjacent regions always have a lower outflow velocity than those measured in the polar coronal holes. Of course, as in all forward modeling methods, the results for the outflow velocities may not be unique.

\placetable{tab_v_uncert}

%%%%%%%%%%%%%%%%%%%%%%%%%%%%%%%%%%
\subsection{Comparisons to IPS Velocity Maps} \label{IPS}

It is interesting to compare our velocity maps with those obtained with IPS observations.  The three most relevant IPS studies that have time periods that overlap with the present work are by \citet{Kojima99, TKF10} and \citet{Breen99}.  The first two references use data from the Solar-Terrestrial Environment Laboratory, Nagoya (STELab) and the last one uses data from the European Incoherent Scatter array (EISCAT). \citet{Kojima99} have IPS velocity data from the period starting from CR~1895 and ending in CR~1917. They project their outflow velocities onto a source surface at $2.5~R_{\sun}$ which is ideal for making comparisons to the velocity structures in our outflow velocity maps. In their Figure 1, they show velocity maps for the whole period, each with contour lines at 300, 350, 400, and $500~km s^{-1}$. The maps are restricted to latitudes between $\pm 30\arcdeg$ because their work emphasized the low speed regions around the streamer belt.  We show in Figure~\ref{fig_comp_ips} the IPS contours only for
$ V < 500~km s^{-1}$ (shown in white) superimposed on our outflow velocity maps for Carrington rotations 1909, 1912, and 1916.  The maps for each rotation in their paper are produced with data from three rotations centered on the primary rotation period.  Figure~\ref{fig_comp_ips}a shows that the IPS low velocity contours follow our low velocity regions (black and dark blue) very well.  In addition, the changes in the structures from one rotation to the next is similar for both data sets.  This comparison provides another confirmation that the slow wind in the heliosphere maps back to the streamer belt during solar minimum as expected.  

\placefigure{fig_comp_ips} 

The major differences between our results and the IPS results in general is the magnitude of the outflow velocities.  The IPS velocities are based on reconstructions which use line of sight data from many elongation angles from the Sun.  The minimum inner boundary of these reconstructions is around 0.2~AU for these full coverage maps \citep{Kojima99}.  Since a constant velocity approximation is used to extrapolate the IPS velocity reconstructions back to the source surface, the IPS velocities at $2.5~R_{\sun}$ will be identical to the velocities at their minimum reconstruction distance.  This appears to be the case for the maps produced by \citet{TKF10} as well.   Their map for the 1996 solar minimum (Figure 1c in their paper) shows "low speed" regions mapping back to the streamer belt and "high speed" regions (mapping back to coronal holes.  Again, while the structure is correct, these outflow velocity values are more typical of 1~AU solar wind speeds than the outflow velocities in the inner corona. 

The EISCAT radio telescope array was used to produce IPS velocity maps during CR~1912-1913 for the First Whole Sun Month \citep[see][]{Breen99,Breen00}.  Because EISCAT operates at the relatively high frequency of 931.5 MHz, it can observe as close as $\sim 15~R_{\sun}$ from the Sun. The disadvantage with this array is that there are fewer radio sources that can be observed at this frequency and so the EISCAT maps have velocity measurements at relatively few locations.  The EISCAT outflow velocities are generally lower than those obtained with the STELab telescope but their velocities are still much larger than our results.  For example, regions of 40 to $100~km~s^{-1}$ on our maps are 200 to $300~km~s^{-1}$ on the EISCAT maps; and likewise regions of 130 to $170~km~s^{-1}$ on our maps are 600 to $>800~km~s^{-1}$. Clearly the solar wind is still accelerating beyond our source surface height.

%==========================================
\section{DISCUSSION} \label{discuss}

The Carrington maps produced for this work are useful for providing constraints on parameters that are used in theoretical models of the corona and the solar wind.  For example, many solar wind models propose some form of mass, momentum, and energy input at the coronal base.  In some cases, these inputs are used as free parameters in the model.  However, with the data presented here, we can provide empirically determined constraints on many of these quantities.   We mention only two here: 1) the solar wind mass flux and 2) the derived expansion factors in the corona.  These two topics are briefly discussed below.

The plots for the particle flux as a function of latitude, which are shown in Figure~\ref{fig_lat_var_all} (panels c and f), contrast to what is measured in the distant heliosphere \citep[see for example the data from Ulysses first full orbit][]{McComas00}.  The 1~AU scaled latitudinal profile of the particle flux is essentially constant (to within the measurement errors) from $30\arcdeg$ latitude all the way to the poles.  On the other hand, our data show an increase from mid-latitudes toward both poles.   This difference is partly explained by the fact that the coronal hole expands super-radially so that low latitudes at Ulysses map back to higher latitudes near the poles at $2.3~R_{\odot}$.  Also the gradients along the field lines in both the density and outflow velocity are different for the high latitude fast solar wind and the low latitude sources of the slow solar wind. The different radial gradients could explain how the different particle fluxes that are determined  in the high and low latitude regions close to the Sun become nearly the same when measured at larger distances.

Realistic models of the solar wind will have to explain the differences between the behavior of the fast and slow solar wind in the regions near the Sun.  To test one such model by \citet{cran07}, we show its results for $V, N$, and $NV$ in the plots on the right-hand-side of Figure~\ref{fig_lat_var_all}.  The model results are plotted with dashed lines in the figure.  Note that this is a model for the open magnetic field regions only and so there is no prediction for the plasma parameters in closed-field regions around the equator.  The model is a self-consistent, single-fluid, wave-driven MHD model of an axisymmetric corona.   It requires relatively few inputs, which include a definition of the Alfven wave spectrum determined by measurements of the magnetic footpoint motions in the photosphere.  It includes a coupled chromosphere model that is heated by acoustic waves and cooled primarily by radiation.  The geometry of the specified magnetic field controls the form of the wave damping that leads to heating and acceleration of the solar wind. 

The qualitative agreement between the data and the model is remarkable considering that no adjustments were made to the model.  There are obvious differences near the streamer boundaries but this is most likely due to line of sight effects that are not included in this axisymmetric model.  There also appears to be some asymmetry in the data with the south pole showing a higher particle flux than the north.  This is most likely a real feature and not an artifact, since it remains present even after averaging over a dozen rotations as shown in Figure~\ref{fig_lat_var_all}c.  We will report on a more complete comparison between this model and our data in a future paper. 

The next topic concerns the so-called expansion factors of the solar wind, which describe the geometric spreading of structures in the solar corona.  This expansion can be measured in two different ways. The first method, which has already been presented, uses the areal expansion factor of the entire coronal hole.  A second method relates to the local expansion factors of magnetic flux tubes embedded in the solar wind.  Ideally the flux tube expansion factors can be determined directly from measurements of the coronal magnetic field, however, such measurements do not routinely exist.  In the past, estimates for the flux tube expansion factors were determined by using estimates of the magnetic field determined by potential field models  \citep[e.g., see][]{Wang1990} or  by MHD models \citep[e.g., see][]{Riley10}. However, for the present work, we will use conservation of mass (or particle) flux arguments to compute the flux tube expansion factors from the coronal base to the source surface.  The particle fluxes are derived from our determinations of density and velocity at the source surface.  In the future, we plan to incorporate a magnetic field model to use with the current maps of plasma parameters.

First we investigate how the outflow velocity in the coronal hole at the north pole varies with the areal expansion factor.  Recall that the areal expansion factor, $f_{A}$ (defined in equation \ref{eqn_fAexp}) is a global expansion factor for the entire coronal hole.  Because $f_{A}$ is only a function of radial height, $r$, there is no dependence on latitude or longitude as is normally used for the magnetic flux tube expansion factors.  We considered using an average outflow velocity for the entire coronal hole but this could possibly dilute any relationship between $f_{A}$ and $V$.  Instead we use the empirical outflow velocities $V(90 \arcdeg )$, determined at the north pole for each Carrington rotation, and plot these as a function of the coronal hole expansion factor. Figure~\ref{fig_vpole_fAexp} shows that outflow velocity at the center of the coronal hole tends to increase with larger expansion factors for the coronal hole. 

\placefigure{fig_vpole_fAexp}

The actual relationship between $V(90 \arcdeg )$ and $f_{A}$ may not have a constant slope for the entire range of expansion factors plotted.  For example, the slope for the data with  $f_{A} \le 3$ may be considerably less steep than the portion of the curve with $f_{A}  > 3$ . However, considering the size of the error bars we will use the same slope for the entire data set in the figure.   Figure~\ref{fig_vpole_fAexp} shows that larger coronal holes have the largest expansion factors and the fastest solar wind.  While this result is not new for solar wind measurements at 1~AU, it is a new result for quantitative velocity measurements made close to the Sun.  An empirical fit to the $V(90 \arcdeg )$ -- $f_{A}$ relationship is shown in the figure. Note that the relationship will have different values for the fitting parameters at a different source surface height used for calculating $V(90 \arcdeg )$ and $f_{A}$.  This fit does not necessarily contradict the idea that {\it flux tube} expansion factors have an {\it inverse} relationship with solar wind speed since, to state again, we are using a {\it global} expansion factor for the entire coronal hole.   

Our results are similar the results of  \citet{nolte76} which showed that the maximum solar wind speed (at 1~AU) had a positive correlation with increasing area of the coronal hole source regions on the disk.  However, there are differences between the two measurements. \citet{nolte76} looked at equatorial coronal holes and they used solar wind speed measurements at 1~AU which could be affected by the transit of the wind through the interplanetary medium.  The data in Figure~\ref{fig_vpole_fAexp} are for a polar coronal hole with velocity measurements made much closer to the Sun at $2.3~R_{\sun}$.  If the coronal hole area on the disk is used, our data give a linear correlation coefficient for a liner fit of V and $A_{CH}$ of only 0.42.  The correlation coefficient using V and the coronal hole area on the source surface has a value of 0.72, which is significantly better. This is another way of saying that while the coronal hole area on the disk is important, the areal expansion factor is a much more important factor in determining the solar wind speed.  The linear correlation coefficient for the fit in Figure~\ref{fig_vpole_fAexp} is 0.83.

We now turn to a discussion of the local (versus global) expansion of the polar coronal hole.
In order to estimate the expansion factors for the local magnetic flux tubes (i.e., the {\it flux tube expansion factors, $f_{exp}$}) we will use mass flux conservation (or actually we use particle flux conservation by dividing the mass flux by the particle mass, $m_p$).  The expression for the particle flux conservation for material flowing from the base of the corona ($r_{o} \approx 1 R_{\sun}$) to the source surface height ($r_{ss} = 2.3 R_{\sun}$) is 
\begin{equation} \label{eqn_fcon}
N_{ss} V_{ss} A_{ss} = N_{o} V_{o} A_{o} , 
\end{equation}
where $N$, $V$, and $A$ are the electron density, outflow velocity, and flux tube area respectively at the source surface (designated with subscript ``ss'') and at the coronal base (with subscript ``o'').  The areas, $A_o$ and $A_{ss}$, are now no longer the entire coronal hole area at the respective heights but are the elemental flux tube areas for the plasma flow.  We now define the traditional expansion factor, $f_{exp}$, which gives the expansion of a flux tube in going from the coronal base to the source surface height, from the following expression:
\begin{equation} \label{eqn_areas}
 \frac{A_{ss}}{A_{o}}  = f_{exp}  \frac{{r_{ss}}^2}{{r_{o}}^2} .
\end{equation}
Combining equations (\ref{eqn_fcon}) and (\ref{eqn_areas}) and rearranging terms gives the final expression for the expansion factor at the source surface height:
\begin{equation} \label{eqn_fexp}
f_{exp}  =  \frac{N_{o} V_{o} }{ N_{ss} V_{ss}{r_{ss}}^2 } .
\end{equation}
As expected, we now see that the outflow velocity at the source surface, $V_{ss}$, is inversely related to the flux tube expansion factor, $f_{exp}$.

We can compute values for the particle flux ($N_{ss}V_{ss}$) for any element on our selected source surface height by using the independent determinations of density and outflow velocity for each Carrington rotation in our data set.  The particle flux at the coronal base ($N_{o}V_{o}$)  can be estimated by assuming that  $f_{exp} = 1$ on the axis of a polar coronal hole and using equation  (\ref{eqn_fexp}).  By taking the mean of the particle flux at the north pole we obtain a value for the particle flux at the coronal base $N_{o}V_{o} = 1.1 \times 10^{13}  ~cm^{-2} s^{-1}$.  We have confidence in this value for the base flux since it is identical to the value found by using the high latitude proton particle flux from Ulysses during its first polar pass at solar minimum \citep{McComas00}. 
We also assume that the base particle flux, $N_{o}V_{o}$, is constant for all latitudes on the Sun where there is open flux (magnetic field lines that make it out to the extended corona).  This last point is a working assumption only but is the most simple argument for the coronal base.   Later on in this section, we will discuss the consequences of relaxing this condition for a constant base particle flux.

In order to understand how the flux tube expansion factors, $f_{exp}$, vary in the corona we could compute the particle fluxes for every spatial bin in each of the Carrington rotation maps.  Variations of the particle flux and expansion factor with latitude can then be found. However we find that it is difficult to determine trends in this way because of the large scatter in the data that tends to mask any correlations.  There is also the problem that the streamer belt is not constant and its spatial locations vary with time for each map.  One solution to these problems is to use density instead of latitude as the independent parameter.  This choice has the advantage that the density is a more suitable physical parameter for describing different coronal structures.  Since the electron density changes monotonically from the pole to the streamer belt/current sheet, it can be used as a proxy for different types of coronal structures regardless of their actual map locations.   In Figure~\ref{fig_pflux_fexp1}a, we show particle fluxes at the source surface as a function of electron density for CR 1909 to 1922.  This period was chosen because the size of the north coronal hole was relatively stable during this time.  The data points are mean values for the discrete density bins for each Carrington rotation map.  The density bins are determined by finding the minimum and maximum densities for each map and dividing this range into 10 intervals.  The lowest density bin has a fixed range from 0 to $2 \times 10^5 ~cm^{-3}$ while the remaining 9 bins are equally divided from $2 \times 10^5 ~cm^{-3}$ to the upper density limit.  The density intervals for each Carrington rotation are slightly different because the upper densities are not identical for each map.  Once the density intervals are calculated for each Carrington rotation map, the mean velocities and particle fluxes are then calculated using the data for the same spatial bins in each density interval.   

Figure~\ref{fig_pflux_fexp1}a is a scatter plot for the variation of the particle flux with density at the source surface height.  Even though there are some outlier points, there is clearly a downward trend in the data, i.e., the particle flux decreases toward the high density streamer belt.  Despite the relatively large scatter and a standard deviation of 40\% about the mean particle fluxes, there is still a statistically significant difference between the particle fluxes in the coronal holes and in the high density streamer regions.   The particle fluxes in the coronal hole at $2.3~R_{\sun}$ are at least two times higher than the particle fluxes from the high density streamers at the same height.  This is in contrast to the 1 AU particle fluxes measured using other techniques which suggest a higher particle (mass) flux over the poles \citep[e.g., see][]{Quem06, Quem07}.  The differences are most likely due to the assumed solar wind acceleration and the expansion factors for the magnetic flux tubes above our source surface which we do not address here.  The topic of the solar wind expansion from the source surface to 1 AU will be addressed in the future.

The spatial bins located in the coronal hole regions have a density which is less than $2 \times 10^5 ~cm^{-3}$ and this cut-off is indicated in the figure by a vertical mark labeled ``CH''.  Note again that we have excluded the data from closed field regions in the center of the streamer belt, where the outflow velocities are essentially zero.  The remaining high density regions include the edges of streamers where some solar wind outflow is still detected as well as the quiet regions between the polar coronal hole boundary and the streamer belt.  Panel (b) of Figure~\ref{fig_pflux_fexp1} shows the flux tube expansion factors plotted as a function of density for the same time period.  The flux tube expansion factors, $f_{exp}$, are calculated by using equation (\ref{eqn_fexp}).  Note how  $f_{exp}$ is relatively constant at low densities but then it increases for density values above $6 \times 10^5 ~cm^{-3}$.  We believe that this is a real effect since the uncertainty in $f_{exp}$ is only $\sim \pm 1$ or less.

\placefigure{fig_pflux_fexp1}

\placefigure{fig_pflux_fexp2}

The particle fluxes, $NV$, and expansion factors, $f_{exp}$, on the source surface have also been computed for later Carrington rotations, i.e., CR~1923 through CR~1936.  These data are plotted in Figure~\ref{fig_pflux_fexp2}.  There are some similarities as well as differences when these plots are compared to the plots of the earlier Carrington rotations in Figure \ref{fig_pflux_fexp1}.
Close examination of data in the two figures shows that both the particle flux and expansion factors are nearly the same for the low density coronal hole regions in both time periods.  This indicates that the density and outflow velocity in the coronal holes compensate for each other for the entire data period from CR~1909 -- 1936.   The same is true for the surrounding quiet regions (i.e., densities between $2 - 4 \times 10^5 ~cm^{-3}$), where there is very little difference between the earlier and later Carrington rotations. However, for the higher density (e.g., $N > 5 \times 10^5 ~cm^{-3}$) regions surrounding the streamer belt there is a much greater sensitivity of the expansion factors to changes in the electron density during the earlier time period.  Figure \ref{fig_pflux_fexp2}a shows a trend with higher expansion factors at higher electron densities. This differs significantly from the trend in Figure \ref{fig_pflux_fexp2}b which shows a tight dependence of $f_{exp}$ with coronal density.  In other words, during CR~1923 -- 1936,  $f_{exp}$ is essentially independent of N,  with a value for $f_{exp}$ slightly greater than 1.

The differences between the expansion factors in Figures \ref{fig_pflux_fexp1} and
\ref{fig_pflux_fexp2} suggest that the large expansion factors at the edges of streamers,
as suggested by \citet{Wang1990} and others, may exist only at solar minimum.  This is the
period with the largest coronal holes and a high density streamer belt confined to a very
narrow latitude range.  At later times in the solar cycle there is no evidence of such
large expansion factors using our data.   It should be noted that the \citet{Wang1990}
expansion factors are calculated differently since they use extrapolated values of the coronal magnetic field at the source surface.  Also, care must be used in comparing numerical values by different authors since some researchers calculate $f_{exp}$ by evaluating the expansion from some lower height to 1~AU, as opposed to our approach of calculating the expansion from the base of the corona to the source surface. 

The expansion factors in our study have a maximum value of about 3 to 4 times an $r^2$
expansion from the coronal base to  $2.3~R_{\odot}$.  Some coronal models have even higher
expansion factors near the coronal hole/streamer edges e.g., \citet{cran07}.  These higher
expansion factors can be accommodated if, for example, we relax the choice for a constant
particle flux at the corona base.  If $N_{o}V_{o}$ is not constant but instead increases
from the coronal hole axis to the streamer belt, then the expansion factors near the streamer belt can be larger than the factor of four shown by our data.  For example, an increase of the base particle flux near the streamer boundary by a factor of two over the coronal hole value would make the expansion factor $f_{exp} \sim 8$ near the streamer boundary.

Another interesting result of this study is the fact that $f_{exp}$ is nearly constant for
most of the corona with values between 1 and 2. This suggests that there is very little
super-radial expansion inside of coronal holes and the surrounding quiet regions (not
including the boundaries of the streamer belt).  A possible explanation for this is that
there may be two different types of slow solar wind as suggested by \citet{Abbo10}.  Deep in the solar minimum period,
the slow wind is produced by the super-radial expansion of the magnetic flux tubes anchored in the corona.  Later on in the solar cycle when $f_{exp}$ is nearly the same everywhere, the slow wind is probably a result of the fact that the solar wind source regions are denser and this therefore requires more energy to accelerate the wind to the same speed as in the less dense coronal holes.

A different way to show the relationships between the outflow velocity, density, and particle flux is shown in Figure~\ref{fig_vel_den}. Once again we show that the outflow velocity structure of the corona at the deepest part of solar minimum (when the coronal holes are the largest) is indeed different from the more radial structure that is present when the corona is in its rising phase.  Panel (b) shows that the data from later in the 
solar minimum period (CR 1923 -- 1936) can
be fit using a simple equation of the form  $N_{ss}V_{ss} = 1.9 \times 10^{12} ~cm^{-2}
s^{-1} $.  The fit works because the $f_{exp}$ is approximately constant for all densities
during this period.  The data from deep in solar minimum (panel a) do not follow the
same fit (solid curve).  A better fit is the dashed curve which is the equation
$N_{ss}V_{ss} = 1.5 \times 10^{12} ~cm^{-2} s^{-1} $.   However, neither fit is
particularly good especially at densities above $\sim 6 \times 10^{5} ~cm^{-3}$.  The
outflow velocities corresponding to the spatial bins with these densities are much lower
than either fit, which is another way of showing that the expansion factors $f_{exp}$
become larger with increasing density during the deepest part of Solar Minimum.  Since we
know that the highest densities correspond to regions in the streamer belt, the best
explanation is that there is a super-radial expansion of the solar wind flow near the boundary of the streamer belt.

\placefigure{fig_vel_den}

%==========================================
% 
% Conclusions
\section{Conclusions}\label{concl}
 
This is the first paper on our efforts to produce complete outflow velocity maps in the corona using over a decade of UVCS and LASCO observations of the corona.  We use the data from these observations along with the well established Doppler dimming analysis of UV spectral lines to make the outflow velocity maps.   We present maps for both outflow velocity and density from the period of low solar activity at the start of cycle 23.  This period (Carrington Rotations 1909 -- 1937) was selected since it shows a clear distinction between the fast and slow solar wind source regions.  

The outflow velocity and density maps are used to reveal new information about the
expansion of the solar wind from the coronal base to our selected source surface height of
$2.3~R_{\odot}$.  We find the following four main results using the data from this work:  
\begin{enumerate}
\item The
boundary of the large polar coronal hole that was studied has a well defined signature
using an outflow velocity threshold that agrees well with estimates of the boundary based
on a density threshold.
\item Our results are consistent with the expected super-radial
expansion with height for the area of polar coronal holes during most of the solar minimum
period, however, there are periods when the areal expansion factor $f_{A}$ is $\approx 1$
in the rising phase of the solar cycle.
\item The flux tube expansion factors $f_{exp}$
(this is different from $f_{A}$) at the source surface height have values between 1 and 2
in the interior of coronal holes.  However, the regions of increased densities near the
streamer belt have a maximum value of $f_{exp} \approx 4$ at  $2.3~R_{\odot}$. The larger
expansion factors may explain the decrease in the solar wind speed at the streamer belt,
however it is found that these large factors are not present when the overall size of the
polar coronal holes start to shrink during the rising phase of the solar cycle.  This
suggests that the expansion factors may not be the only controlling factor that governs the production of the slow speed wind.  A constant particle flux coupled with an increased particle density can produce the same effect of lowering the wind speed.
\item Finally, the comparison of our outflow velocity maps with those derived from IPS measurements show that while the IPS maps have the same general structure of the velocity features as our maps, the absolute magnitudes of the IPS outflow velocities are always much larger than the velocities determined by Doppler dimming. This is true in both coronal holes and streamer regions. We conclude from this that there is additional acceleration of both the fast and slow solar wind beyond $2.3~R_{\sun}$.
\end{enumerate}

What is new about this work is that the expansion factors are calculated using only the
outflow velocity and density measurements made in the extended corona. Our approach is
different from studies of coronal expansion factors that use magnetic field measurements
at the base or at 1 AU, with models to extrapolate the magnetic field values for the regions in between.   In the future it would be useful to compare the results for the expansion factors presented here with results from these models.  

\begin{figure}
\includegraphics[scale=0.65]{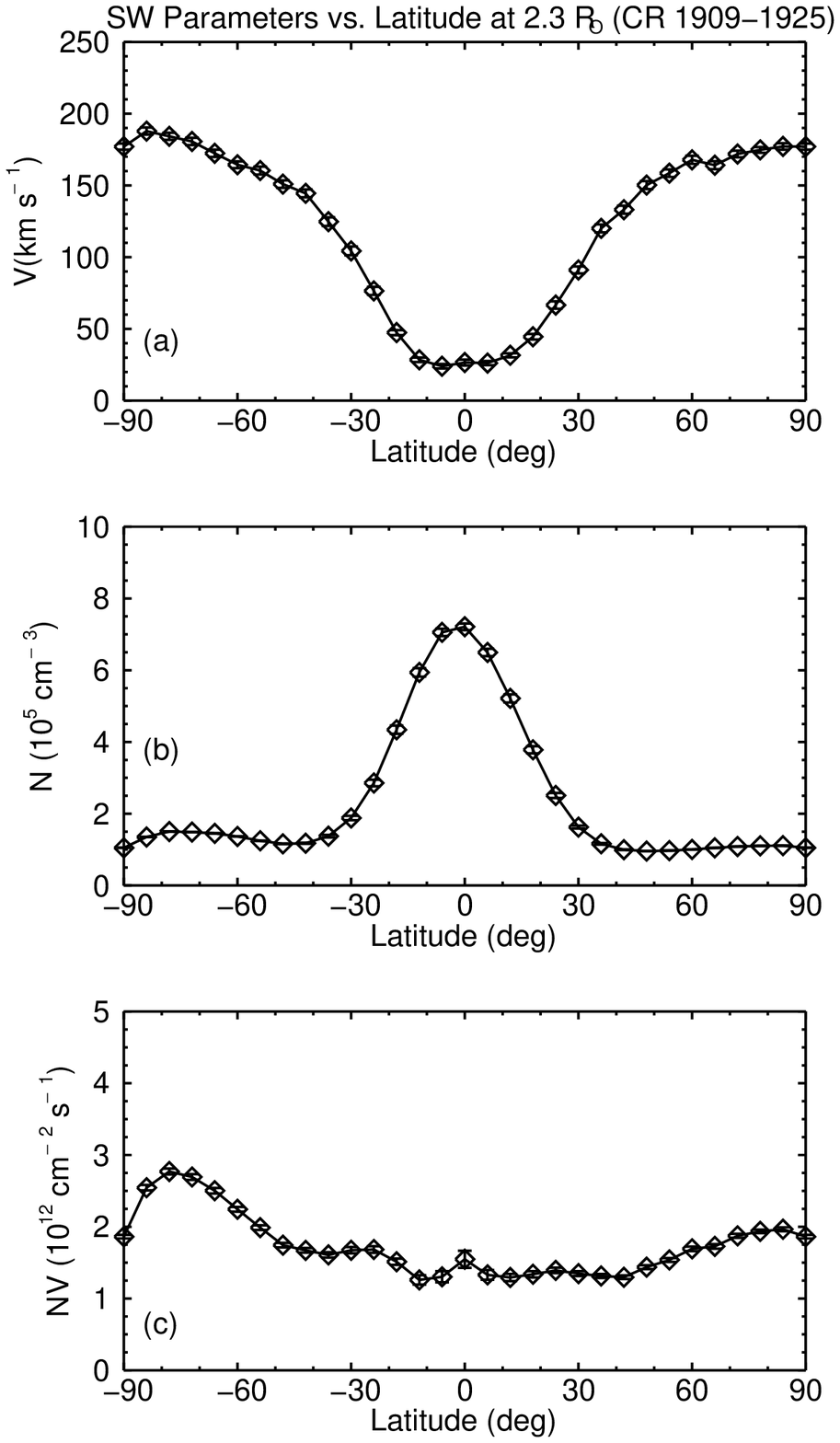}
\includegraphics[scale=0.65]{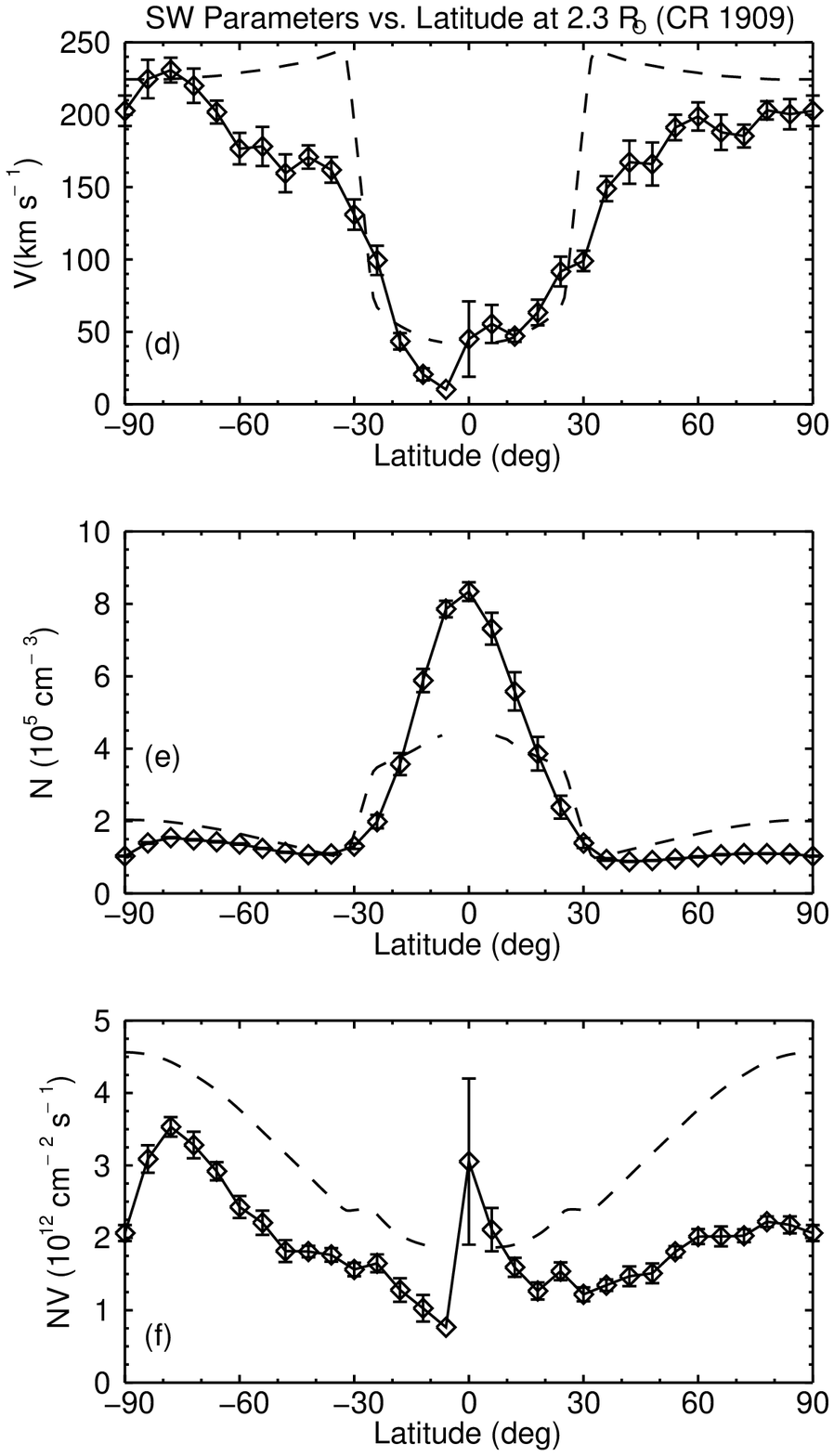}
\caption{On the left are latitudinal profiles that are derived by averaging data in the maps for CR 1909 -- 1925.  Panels (a) and (b) show the averages for the outflow velocity and electron density, respectively, for 30 latitude bins spanning from $-90 \arcdeg$ to $90 \arcdeg$.  Panel (c) shows the average solar wind particle flux (NV), which is approximately constant for all latitudes.
On the right are the same quantities but for the single Carrington Rotation 1909.  Shown for comparison are the results (dashed lines) from an independent theoretical solar wind model by \citet{cran07}.  See the text for more details.
 \label{fig_lat_var_all}}
\end{figure}

\begin{figure}
%%%%%% don't use \plotone here
\includegraphics[scale=0.60]{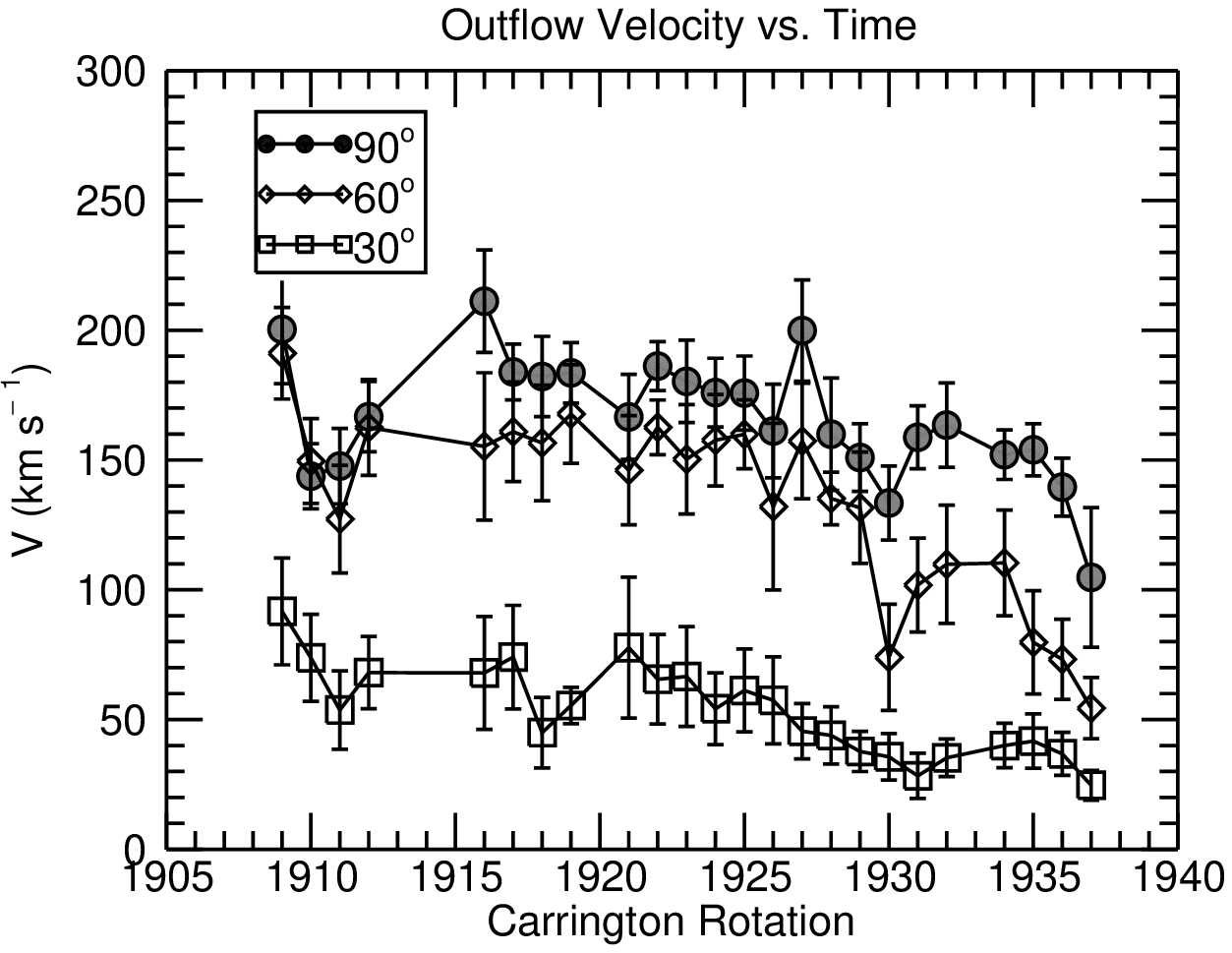}

\includegraphics[scale=0.60]{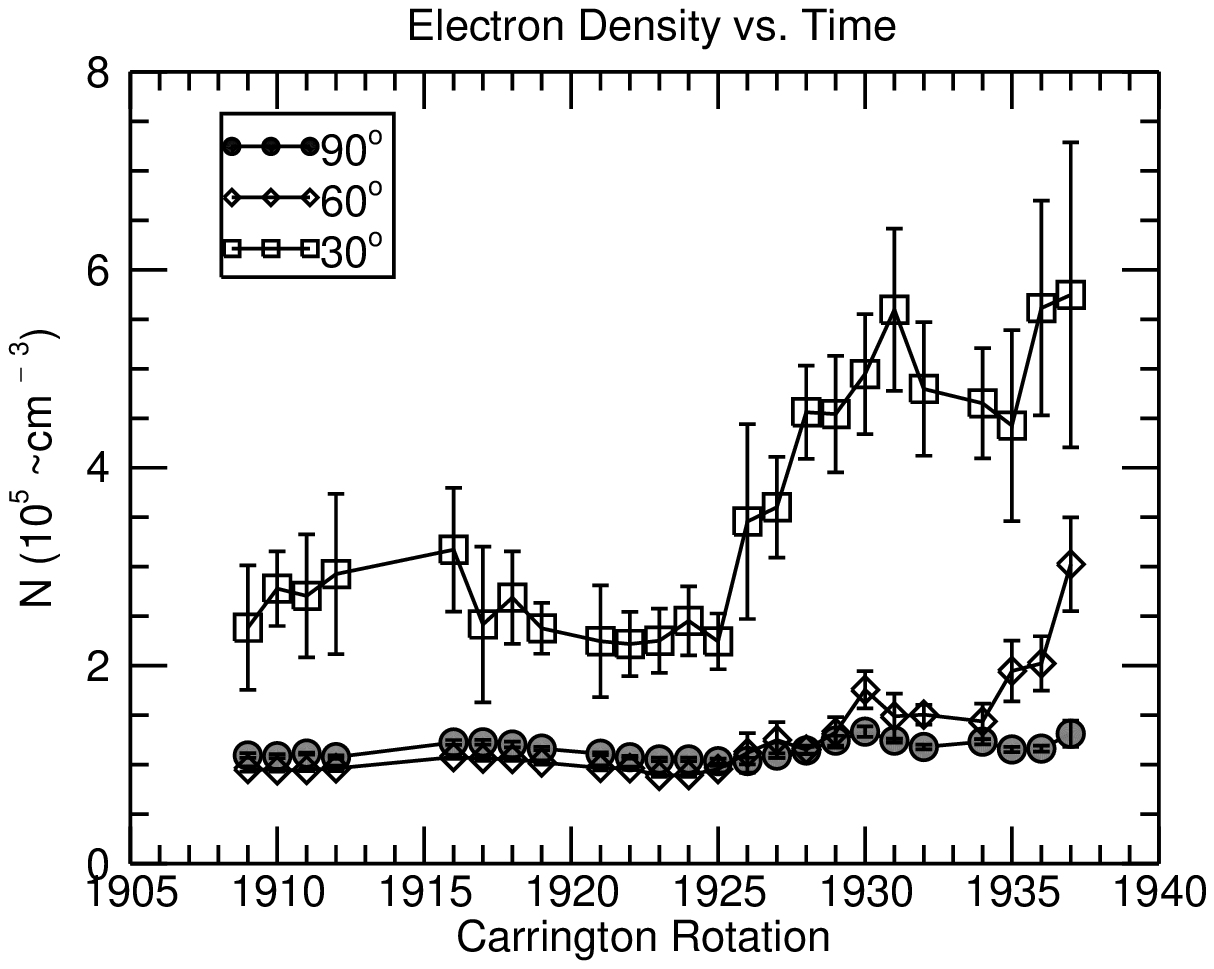}

\includegraphics[scale=0.60]{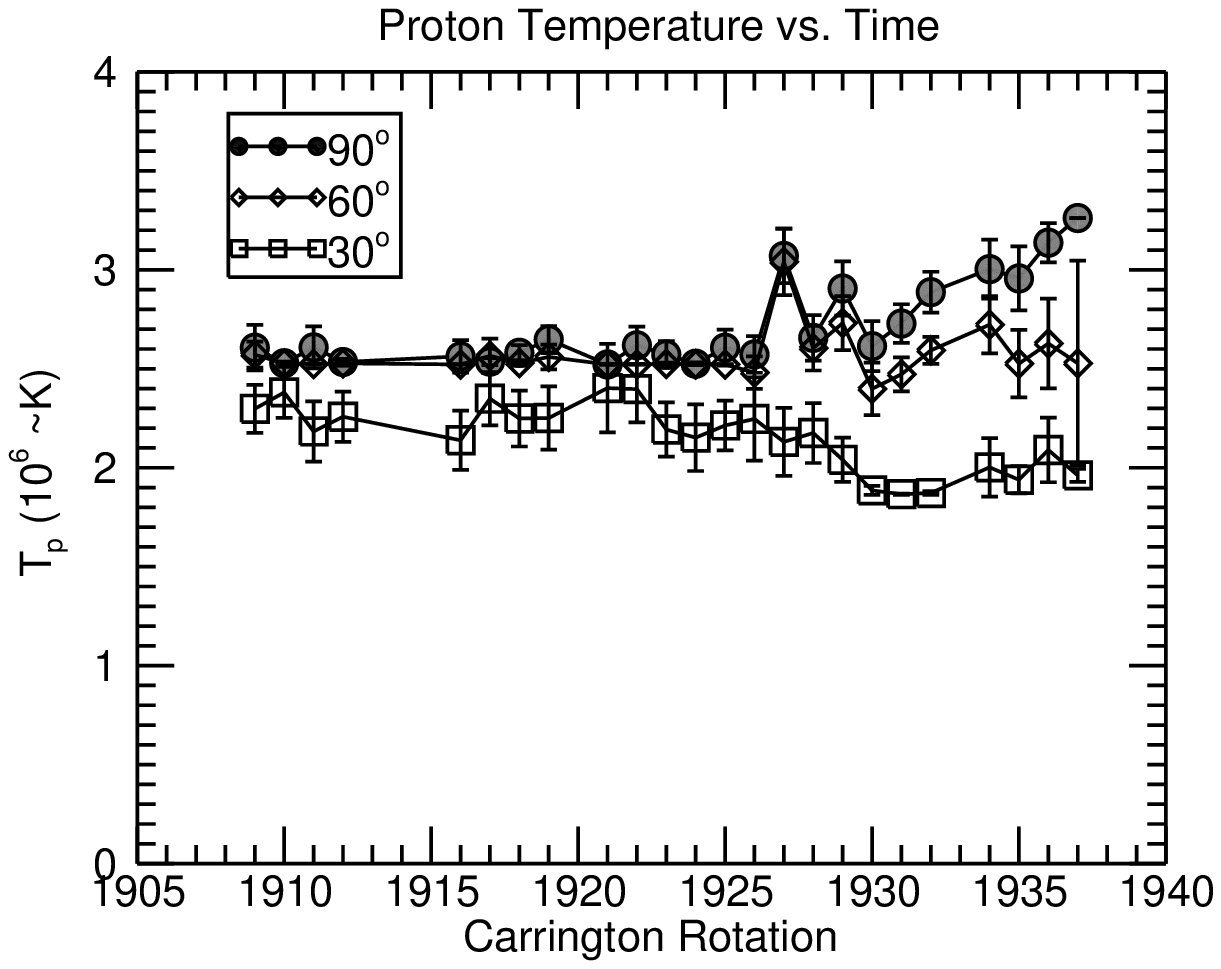}
\caption{Evolution of coronal parameters at $90 \arcdeg$, $60 \arcdeg$, and $30 \arcdeg$ for
outflow velocity (top), electron density (middle), and proton kinetic temperature (bottom). The data are averaged over each rotation with $\pm 2 \sigma$ error bars shown for each data point.  Missing data points are described in the text.
\label{fig_v_3lat_crn}}
\end{figure}

\begin{figure}
\plotone{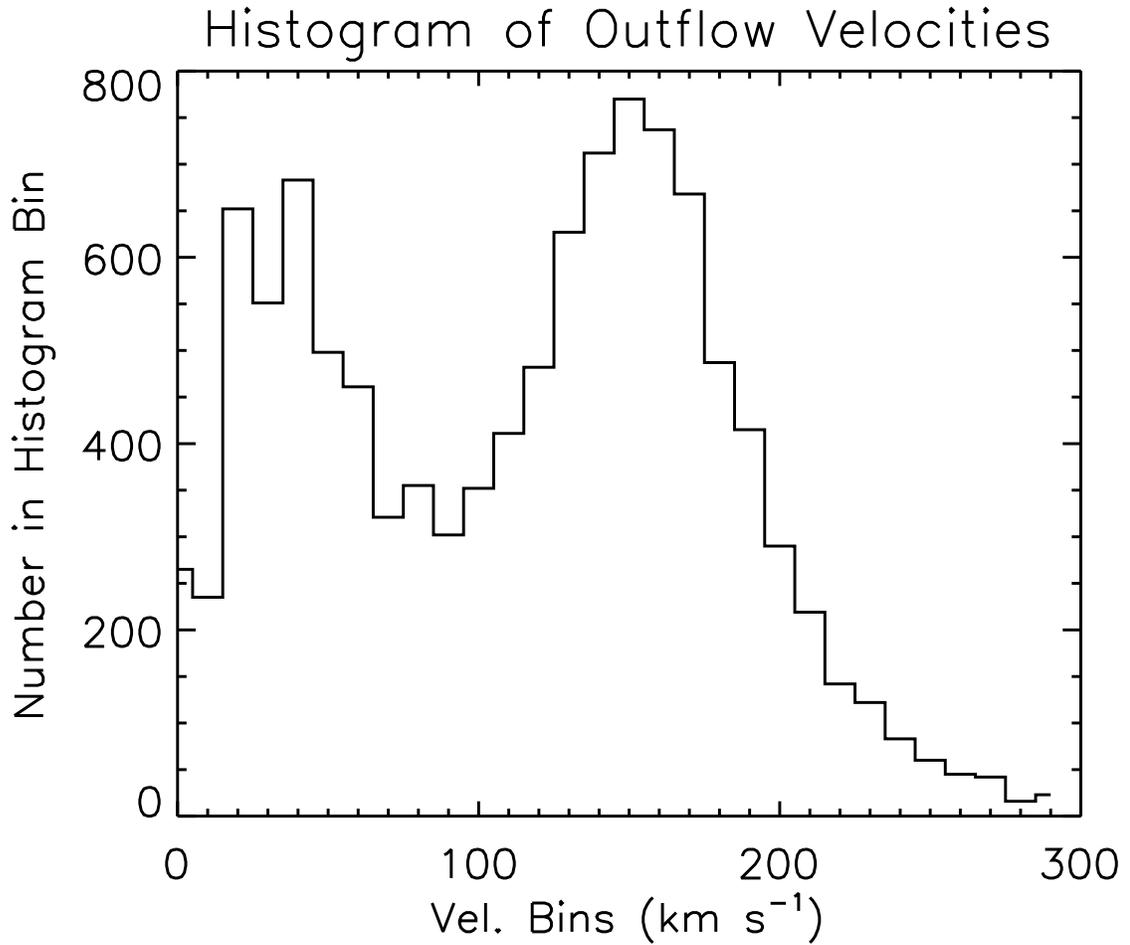}
\caption{Histogram of the outflow velocities computed for all of the latitude/longitude
bins from the maps presented in Figure~\ref{fig_vel_all}.  The bins that contain missing
data or where the outflow velocities were not defined are excluded.
\label{fig_histo_vel}}
\end{figure}

\begin{figure}
\plotone{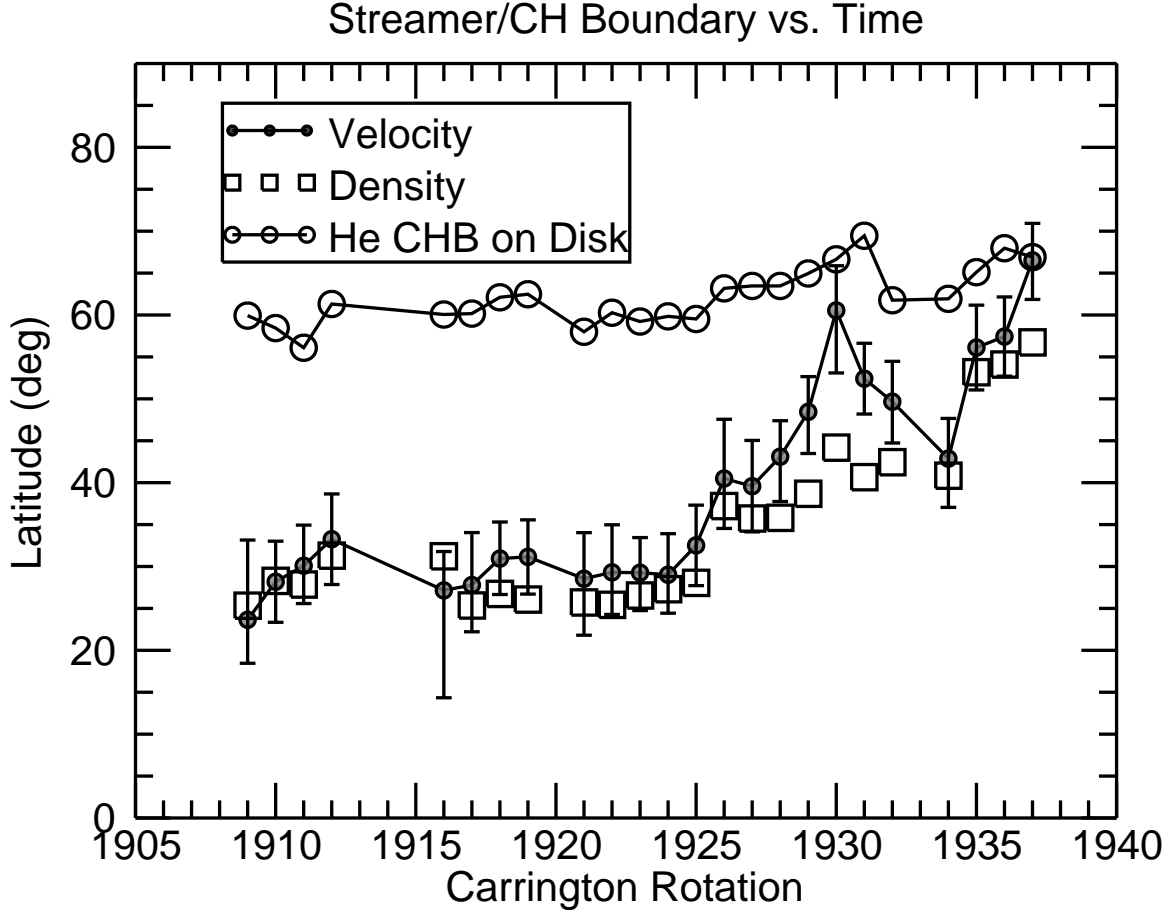}
\caption{Coronal hole boundaries (CHBs) determined from three different methods.  The solid circles are the mean latitude boundary for each Carrington rotation determined using the $100 ~km~s^{-1}$ threshold for the outflow velocities. The open square symbols are the mean coronal boundaries computed by determining the latitude where the electron density is 15 \% of its peak value at the current sheet.  Both are determined at the $2.3~R_{\sun}$ source surface.   The open circles are the mean latitudes for the coronal hole boundaries determined at $ \sim 1~R_{\sun}$ by using observations of chromospheric He~I 1083.0 nm emission \citep{har02}.   See main text for details and the explanation for missing data.
\label{fig_nch_lat_crn}}
\end{figure}

\begin{figure}
\plotone{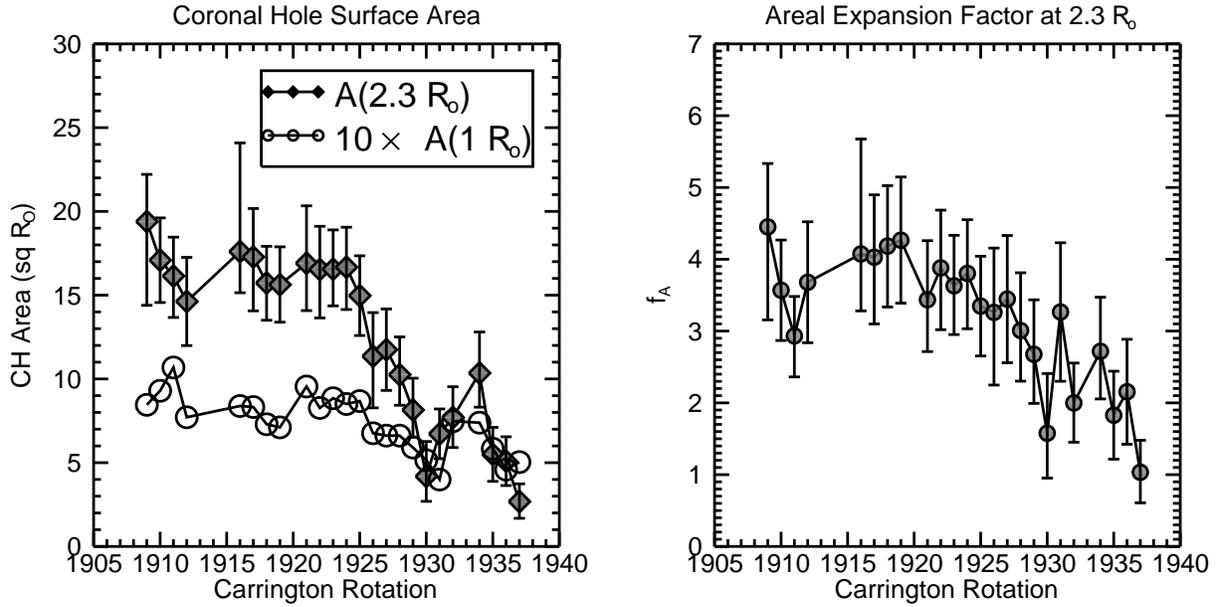}
\caption{{\it Left}:  Coronal hole surface area $A_{CH}(r)$ at $2.3~R_{\sun}$ (solid diamonds) and at the coronal base $r \approx 1~R_{\sun}$ (open circles) as function of time.  The values shown are averaged over each Carrington rotation. {\it Right}:  The coronal hole areal expansion factor $f_{A}$ (defined in equation \ref{eqn_fAexp}) as a function of time.  Both plots show $1 \sigma$ error bars.
\label{fig_ch_area_fexp}}
\end{figure}

\begin{figure}
\includegraphics[scale=0.45]{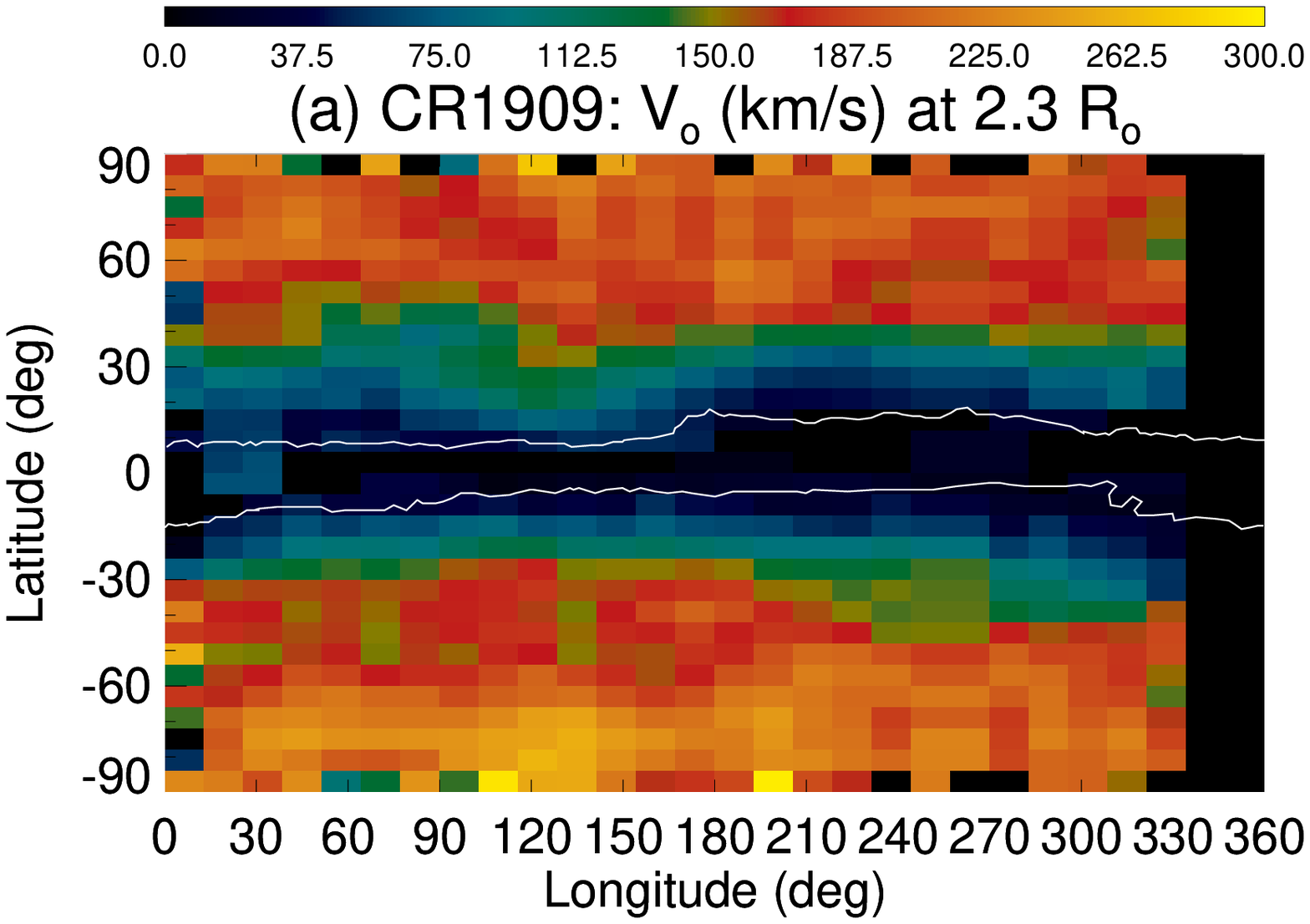}

\includegraphics[scale=0.45]{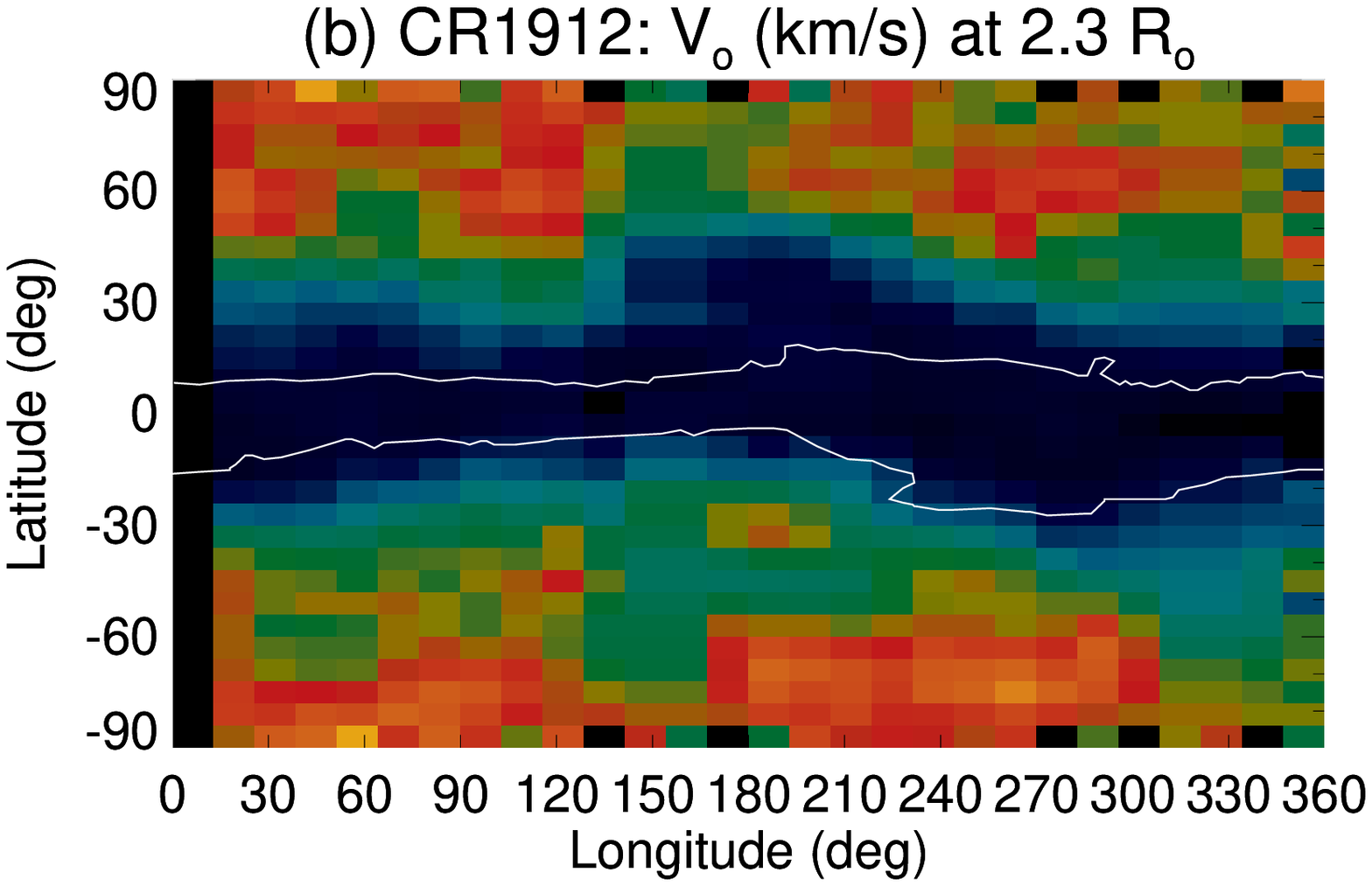}

\includegraphics[scale=0.45]{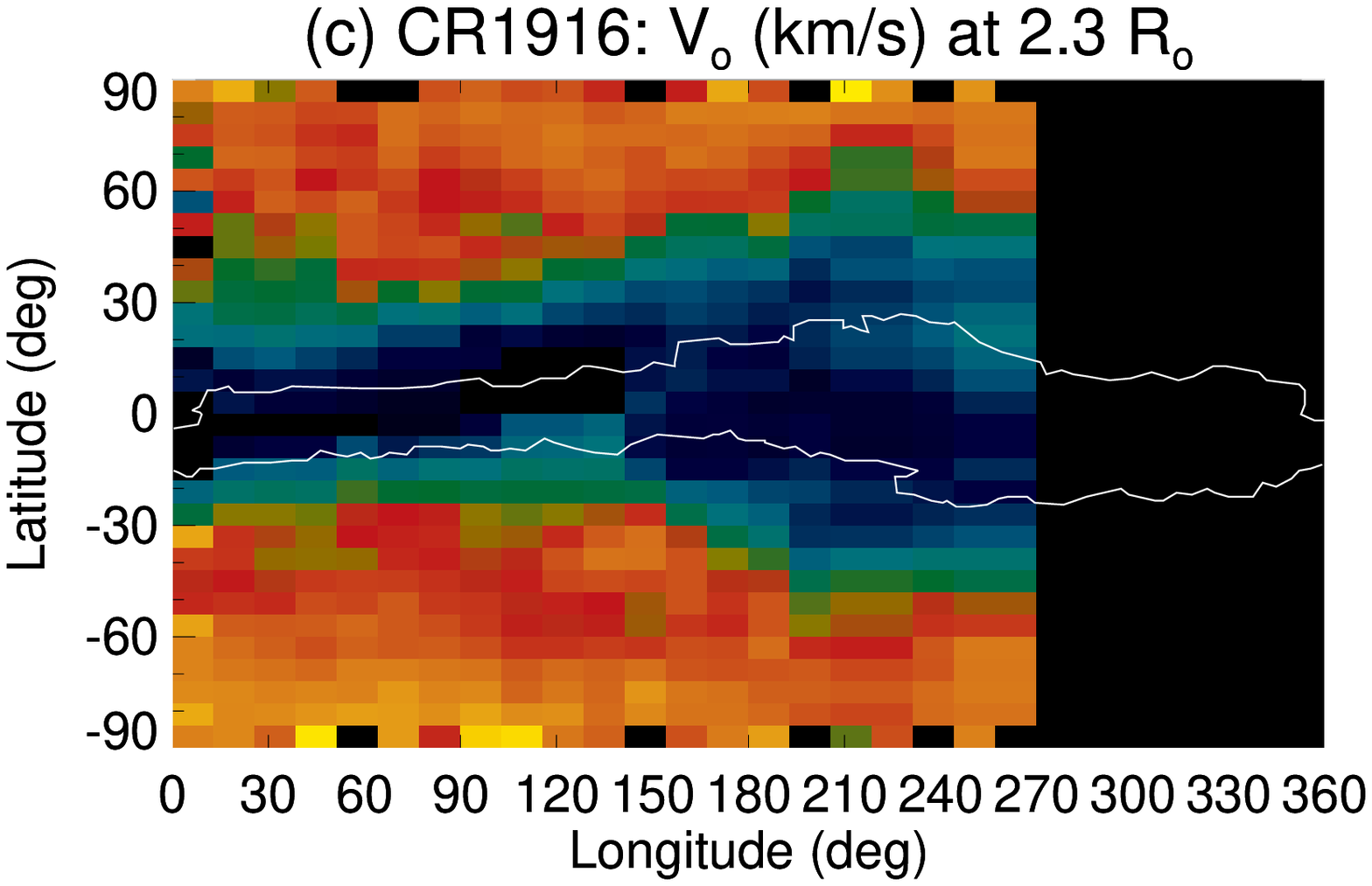}
\caption{Outflow velocity maps for CR~1909, 1912, and 1916 which span the period included in the overlap of the UVCS and STELab IPS observations for the solar minimum period.  The IPS contours for outflow velocities of $500~km~s^{-1}$ (projected back to $2.5~R_{\sun}$) are shown as a white curve on each map. Regions outside of the boundary have IPS speeds above $500~km~s^{-1}$ and regions inside the boundary have speeds below this value.  The UVCS map for CR~1916 (panel c) is incomplete.
\label{fig_comp_ips}}
\end{figure}

\begin{figure}
\plotone{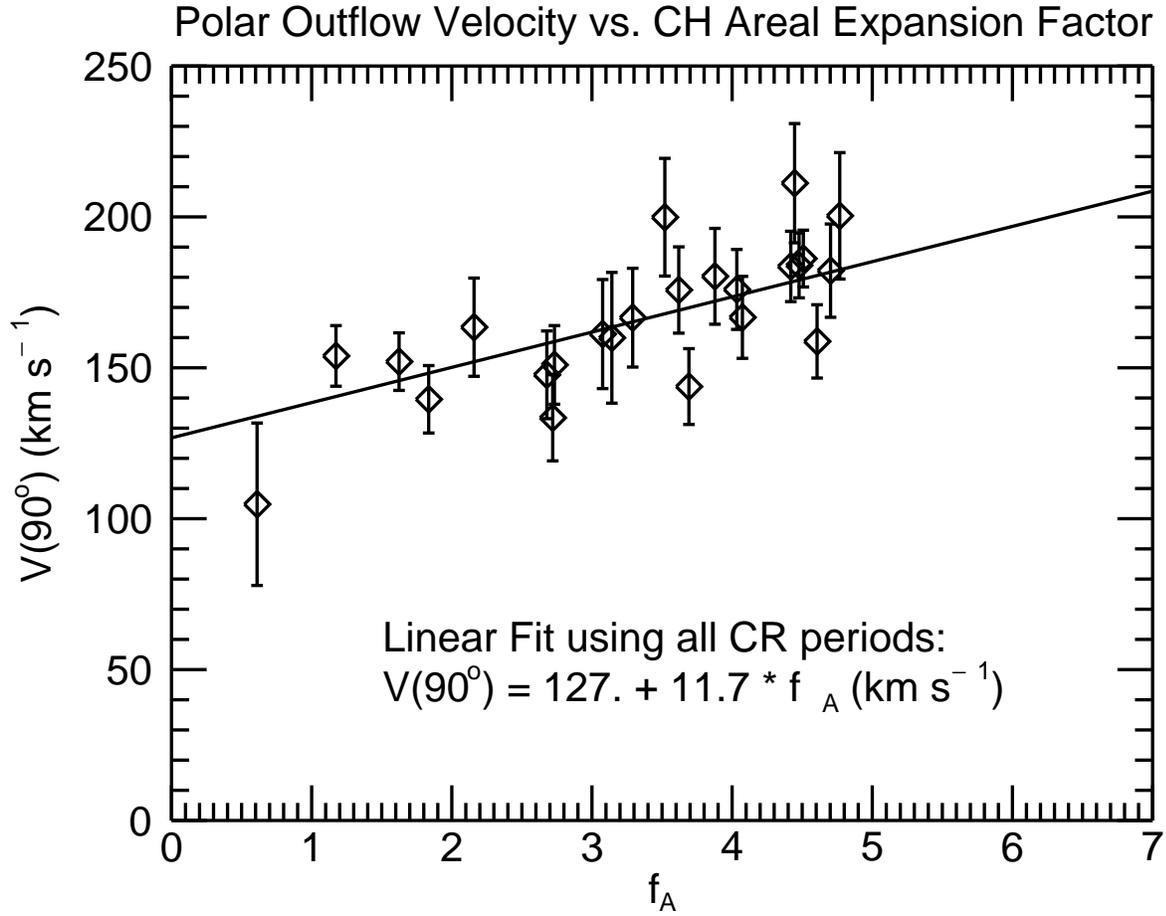}
\caption{Outflow velocity over the north pole, $V(90 \arcdeg )$, vs. coronal hole expansion factor, $f_{A}$, at the source surface height of $2.3~R_{\odot}$.  Each data point is the mean for one Carrington rotation with $2 \sigma$ error bars shown.  A linear fit of the data is drawn as a solid line with its equation shown at the bottom of the plot.
\label{fig_vpole_fAexp}}
\end{figure}

\begin{figure}
\plotone{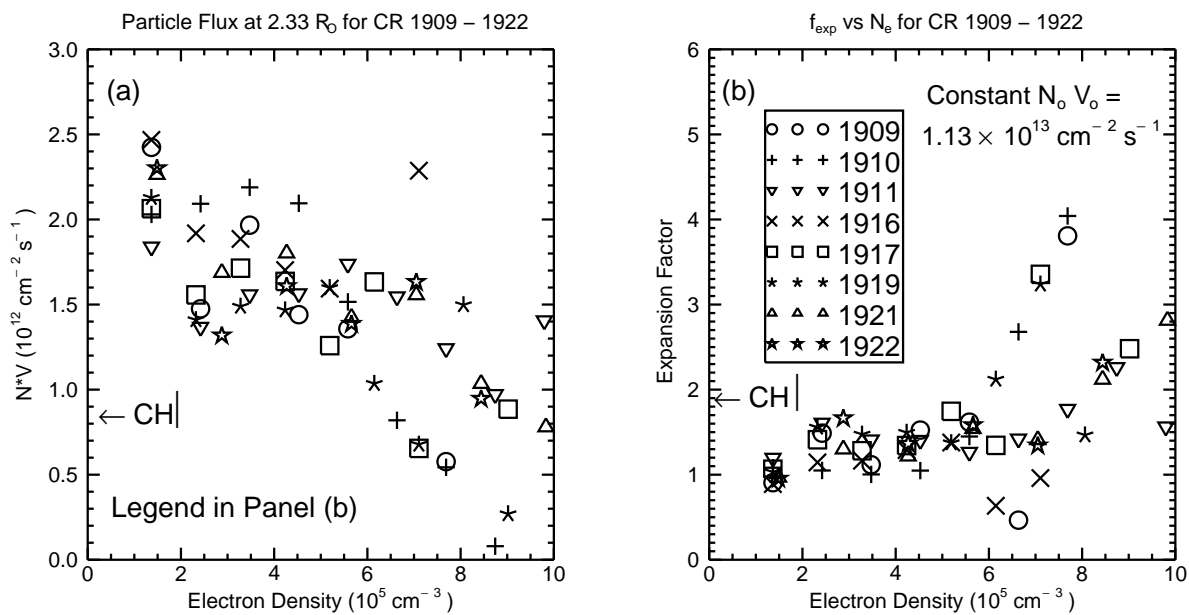}
\caption{(a) Particle flux vs. density at the source surface height of $2.3~R_{\odot}$ for Carrington Rotations 1909 to 1922.  The legend shows the symbols used for the different Carrington Rotation periods. (b) Flux tube expansion factors for the same data in panel (a).  Expansion factors $f_{exp}$ are calculated assuming a constant base flux $N_{o}V_{o} = 1.1 \times 10^{13}  ~cm^{-2} s^{-1}$.
\label{fig_pflux_fexp1}}
\end{figure}

\begin{figure}
\plotone{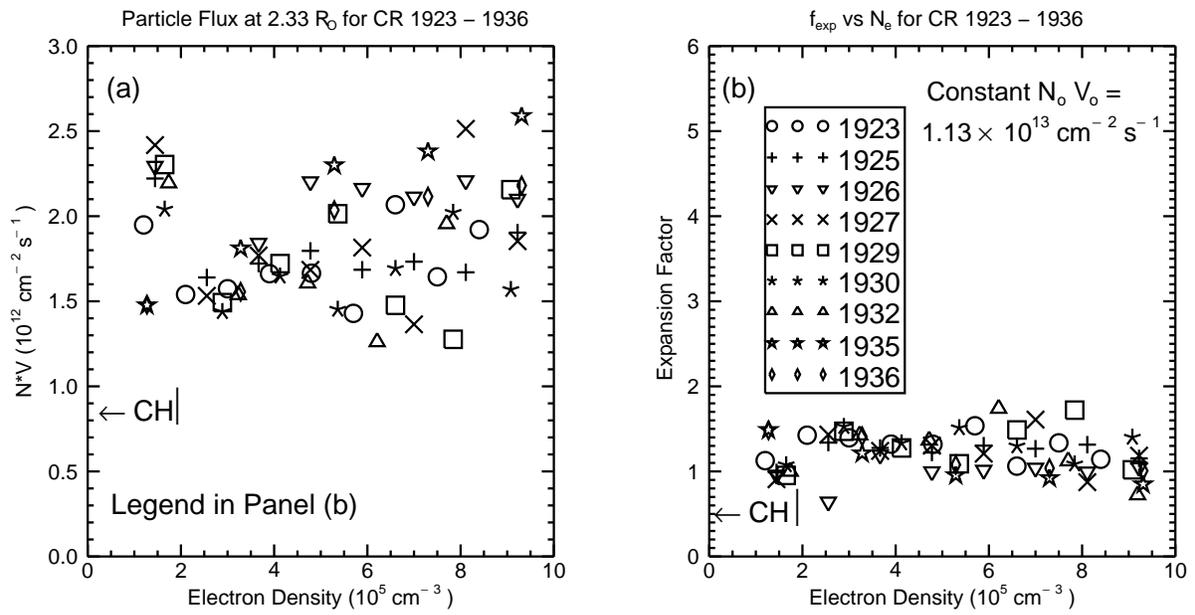}
\caption{Similar data as in Figure \ref{fig_pflux_fexp1} but for Carrington rotations 1923 -- 1936. (a) Particle flux vs. density (b) Flux tube expansion factor vs. density.
\label{fig_pflux_fexp2}}
\end{figure}

\begin{figure}
\plotone{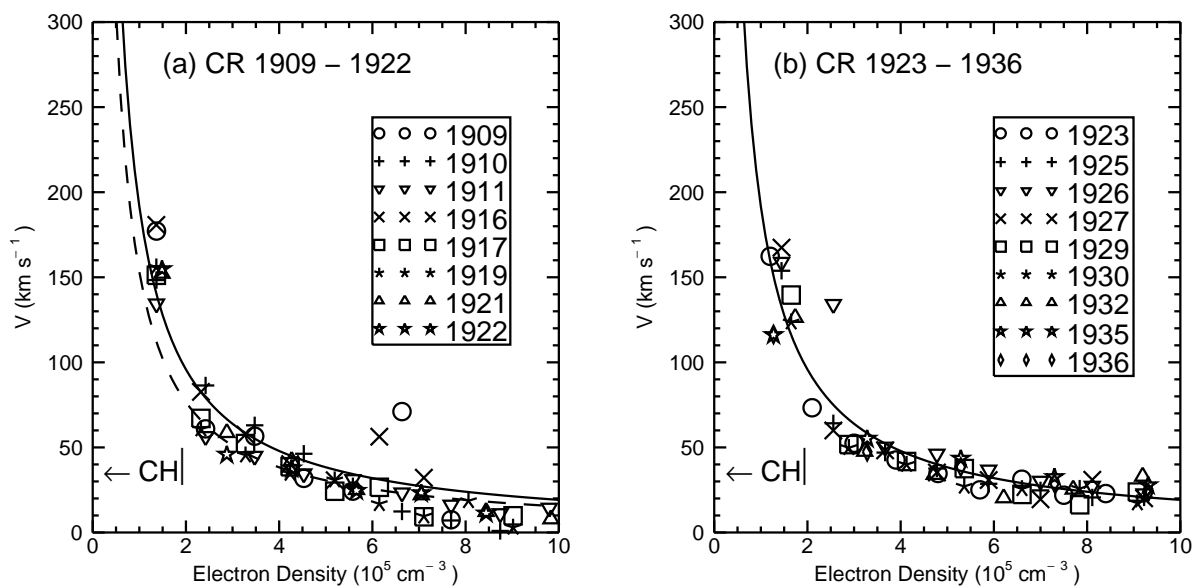}
\caption{Plot of outflow velocity vs. electron density for (a) CR 1909 -- 1922 and  (b) CR 1923 -- 1936.  The dashed curve is the best fit for the data in Panel (a) using $N_{ss}V_{ss} = 1.5 \times 10^{12} ~cm^{-2} s^{-1} $. The solid curve, which is identical in both plots, is the best fit for the data in Panel (b) using the fit $N_{ss}V_{ss} = 1.9 \times 10^{12} ~cm^{-2} s^{-1} $ . Both curves assume $f_{exp}$ is constant for all density intervals.
\label{fig_vel_den}}
\end{figure}

The long term objective for this work is to produce
constraints for coronal and solar wind models that are as free as possible from
assumptions and thus are tied closely to the observational data.  One way to do this is to
use observations made in the solar wind source regions so that there is no need to use
extrapolations from more distant regions in the heliosphere.  Another important aspect of
this work is to make the case that the results of theoretical models to be tested should
not be included in the determination of the physical parameters that are to be used as the
constraints, or vice versa.   With the production of the outflow velocity and density maps
for this work we have accomplished the first step.  The next step is to include these maps
in the ongoing detailed comparisons between coronal observations and the latest
coronal/solar wind models.   We anticipate both our own independent comparisons of our data
with models as well as tests performed by the modeling groups themselves. As a first example, we made a preliminary comparison of our data with a model (in Figure~\ref{fig_lat_var_all}) to show that extended heating and momentum deposition from Alfven waves may be important. A more rigorous quantitative comparison will be reported in a future paper.  Also, for future studies we would like to include the entire period of the UVCS and LASCO synoptic data set, which includes the solar maximum and the declining phases of cycle 23.

%==========================================

\acknowledgments

This work was supported in part by NASA grants 
NNG06GE74G, NNX07AB98G and NNX08AQ96G to the
Smithsonian Astrophysical Observatory.  The LASCO-C2 project at the Laboratoire d'Astrophysique de Marseille (formerly Laboratoire d'Astronomie Spatiale) is funded by the Centre National d’Etudes Spatiales (CNES).  UVCS and LASCO are part of {\it SOHO}, which is a project of international cooperation between ESA and NASA. The authors acknowledge the use of sunspot data from the World Data Center for the Sunspot Index, Royal Observatory of Belgium; and coronal source surface magnetic field data from the Wilcox Solar Observatory Web site at 
http://wso.stanford.edu/synsourcel.html. We also thank S. R. Cranmer for supplying the results of his theoretical model for comparing to our data and an anonymous referee for providing useful comments on the manuscript.

%%%%%%%%%%%%%%%%%%%%%%%%%%%%%%%%%%%%%%%%%%%%
%%%%%%%    List of Figures and Tables %%%%%%%%%%%%%%%%%%%%%%%%%%%%%%%%%%%%%%%%%%%

 %1% Figure~\ref{fig_ne_v2x2}
 %2% Figure~\ref{fig_vel_all}
 %3% Figure~\ref{fig_lat_var_all}
 %4% Figure~\ref{fig_v_3lat_crn}
 %5% Figure~\ref{fig_histo_vel}
 %6% Figure~\ref{fig_nch_lat_crn}
 %7% Figure~\ref{fig_ch_area_fexp}
            %  REIMOVED %8% Figure~\ref{fig_vmax_min}
 %T1% Table~\ref{tab_nch_params}
            % ADD Table 2
 %T2% Table~{tab_v_uncert}
 %8 % Figure~\ref{fig_comp_ips}  [[new]]
 %9 % Figure~\ref{fig_vpole_fexp}
 %10% Figure~\ref{fig_pflux_fexp1}
 %11 % Figure~\ref{fig_pflux_fexp2}
 %12 % Figure~\ref{fig_vel_den}
%----------------------------------------

%%%%  USE \placetable*{key}  AND \placefigure*{key}
%%%%  TO INDICATE TO EDITOR WHERE TO PLACE FIGURES AND TABLES
%
%%%%%%%%%%%%   See instructions for ApJ submissions  %%%%%%%%%%%%%

%=======================================

%% To help institutions obtain information on the effectiveness of their
%% telescopes, the AAS Journals has created a group of keywords for telescope
%% facilities. 
%% After the acknowledgments section, use the following syntax and the
%% \facility{} macro to list the keywords of facilities used in the research
%% for the paper.  Each keyword will be checked against the master list during
%% copy editing.  Individual instruments can be provided in parentheses,
%% after the keyword, but they will not be verified.

Facilities: \facility{SOHO(UVCS,LASCO)}

%=====================================
% \appendix
%% Appendix material should be preceded with a single \appendix command.
%% There should be a \section command for each appendix. Mark appendix
%% subsections with the same markup you use in the main body of the paper.
%% Each Appendix (indicated with \section) will be lettered A, B, C, etc.
%% The equation counter will reset when it encounters the \appendix
%% command and will number appendix equations (A1), (A2), etc.

% ============================================

%%%%%%%%%%%%%%%%%%%%%%%%%%%%%%%%%%%%
%%%%%%%%%%%%%%%%%%%%%%%%%%%%%%%%%%%%

%%%%%%%%%%%%%%  Place Figures Here   %%%%%%%%%%%%%%%%%%%

%%%%%%%%%%%%%%   Place Tables Here   %%%%%%%%%%%%%%%
%%%%%%%%%%% Table 1  %%%%%%%%%%%
\begin{deluxetable}{l c r r r r r}
\tablecolumns{7}
\tablewidth{0pc}
\tablecaption{Plasma and Geometric Properties at $2.3~R_{\odot}$ vs. Time and Latitude 
  \label{tab_nch_params}}
\tablehead{ 
\colhead{} & \colhead{} & \multicolumn{5}{c}{Mean (and Standard Deviation) for Five Periods} \\ 
\cline{3-7} \\ 
\colhead{Parameter} & \colhead{Latitude} & 
\colhead{CR~1909} & 
\colhead{CR~1925} & 
\colhead{CR~1929} & 
\colhead{CR~1934} & 
\colhead{CR~1936} }
\startdata
             $V (km ~s^{-1} )$ &     $+90^{\circ}$ & 200. ( 10. ) & 176. (  7. ) & 151. (  7. ) & 152. (  5. ) & 140. (  6. ) \\
                               &     $+60^{\circ}$ & 191. (  9. ) & 160. (  7. ) & 132. ( 11. ) & 110. ( 10. ) &  73. (  8. ) \\
                               &     $+30^{\circ}$ &  92. ( 10. ) &  61. (  8. ) &  38. (  4. ) &  40. (  4. ) &  37. (  4. ) \\
\hline\
        $N (10^{5} ~cm^{-3} )$ &     $+90^{\circ}$ & 1.09 (0.01 ) & 1.03 (0.01 ) & 1.24 (0.02 ) & 1.23 (0.01 ) & 1.16 (0.01 ) \\
                               &     $+60^{\circ}$ & 0.96 (0.01 ) & 0.95 (0.02 ) & 1.33 (0.08 ) & 1.44 (0.09 ) & 2.02 (0.14 ) \\
                               &     $+30^{\circ}$ & 2.38 (0.31 ) & 2.25 (0.14 ) & 4.54 (0.29 ) & 4.65 (0.28 ) & 5.61 (0.54 ) \\
\hline\
            $T_p (10^{6} ~K )$ &     $+90^{\circ}$ & 2.61 (0.06 ) & 2.61 (0.05 ) & 2.91 (0.07 ) & 3.00 (0.07 ) & 3.14 (0.05 ) \\
                               &     $+60^{\circ}$ & 2.57 (0.03 ) & 2.52 (0.00 ) & 2.73 (0.07 ) & 2.72 (0.07 ) & 2.63 (0.11 ) \\
                               &     $+30^{\circ}$ & 2.30 (0.06 ) & 2.21 (0.06 ) & 2.04 (0.06 ) & 2.00 (0.07 ) & 2.09 (0.08 ) \\
\hline\
           $\theta_{B} ~(deg)$ &    \nodata & \multicolumn{1}{c}{ 24. (7.4)} & 
\multicolumn{1}{c}{ 33. (4.8)} & \multicolumn{1}{c}{ 48. (4.6)} & \multicolumn{1}{c}{ 43. (5.3)} & \multicolumn{1}{c}{ 57. (4.7)} \\
  $A_{nch} ~({R_{\sun}}^{2} )$ &    \nodata & \multicolumn{1}{c}{19.4 (3.9)} & 
\multicolumn{1}{c}{15.0 (2,4)} & \multicolumn{1}{c}{ 8.1 (1.8)} & \multicolumn{1}{c}{10.3 (2.2)} & \multicolumn{1}{c}{ 5.1 (1.5)} \\
                    $f_{A} $ &    \nodata & \multicolumn{1}{c}{ 4.5 (1.1)} & 
\multicolumn{1}{c}{ 3.3 (0.7)} & \multicolumn{1}{c}{ 2.7 (0.7)} & \multicolumn{1}{c}{ 2.7 (0.7)} & \multicolumn{1}{c}{ 2.2 (0.73)} \\
\enddata 
\tablecomments{ The quantities shown are averaged over an entire rotation with their estimated standard deviations shown within the parentheses. 
The parameter $T_p$ is the proton kinetic temperature that 
includes both thermal and nonthermal (e.g., wave) heating of the protons. }
\end{deluxetable} 

%%%%%%%%%%% Table 2 %%%%%%%%%%%%%%
\begin{deluxetable}{l c r r r}
\tablecolumns{5}
\tablewidth{0pc}
\tablecaption{Estimates for Uncertainties in the Outflow Velocity
  \label{tab_v_uncert}}
\tablehead{ 
\colhead{} & \colhead{} &\multicolumn{3}{c}{Model for $T_{\|}$} \\ 
\cline{3-5} \\ 
\colhead{Outflow Velocity Regime} &  
\colhead{~~} &
\colhead{$T_{\bot}/10$} & 
\colhead{$T_{\bot}/6$} &  
\colhead{$T_{\bot}/1$} }
\startdata
    High speed (polar)       & ~  & $177.~ ^{+23} _{-25}$ & $206.~ ^{+30} _{-37}$ & $292.~ ^{+68} _{-67}$ \\
    Moderate speed (midlat.) & ~ &  $124.~ ^{+11} _{-11}$ & $135.~ ^{+15} _{-16}$ & $169.~ ^{+25} _{-28}$  \\
    Low speed (equatorial)  & ~ & $26.~ ^{+32} _{-25}$  & $26.~ ^{+32} _{-25}$ & $0.~ ^{+60} _{-0}$  \\
\enddata 
\tablecomments{Table values shown are the empirical outflow velocities (in $km ~s^{-1}$) with their $\pm 1 \sigma$ uncertainties based on the uncertainties in the \ion{O}{6} intensity ratios and input plasma parameters.  Entries are for three different temperature anisotropies and for three different latitude bins ($66\arcdeg$, $36\arcdeg$, or $0\arcdeg$) during CR~1912. See the main text for details.}
\end{deluxetable}

%========================================
\end{document}